\documentclass[reprint,amsmath,aps,prb,superscriptaddress]{revtex4-2}
\usepackage{lineno}

\usepackage{graphicx}
\usepackage{float}
\usepackage{epstopdf}
\usepackage{dcolumn}
\usepackage{hyperref}
\hypersetup{
    colorlinks   = true, 
    urlcolor     = blue, 
    linkcolor    = blue, 
    citecolor    = blue 
}
\usepackage{color}
\usepackage{amsmath}
\usepackage{xspace}
\usepackage{times}
\usepackage{color}

\usepackage{booktabs}
\usepackage{lmodern} 

\newcommand{\mycol}{1}
\usepackage{amsmath}
\usepackage[english]{babel}
\usepackage{pythonhighlight}

\newcommand{\DMTTFX}{($o$\,-DMTTF)$_2X$\xspace}
\newcommand{\Cl}{($o$\,-DMTTF)$_2$Cl\xspace}
\newcommand{\Br}{($o$\,-DMTTF)$_2$Br\xspace}
\newcommand{\I}{($o$\,-DMTTF)$_2$I\xspace}
\newcommand{\TSP}{$T_{\text{SP}}$\xspace}
\newcommand{\etal}{\textit{~et al.}\xspace}

\newcommand{\PF}{(TMTTF)$_2$PF$_6$\xspace}

\newcommand{\SI}{
	\setcounter{table}{0}
	\renewcommand{\thetable}{S\arabic{table}}
	\setcounter{figure}{0}
	\renewcommand{\thefigure}{S\arabic{figure}}
	\setcounter{equation}{0}
	\renewcommand{\theequation}{S\arabic{equation}}
}

\definecolor{bluegray}{rgb}{0.4, 0.6, 0.8}
\newcommand{\bluetitle}{\color{bluegray}}

\newcommand{\affIMNP}{CNRS, Aix-Marseille Universit\'{e}, IM2NP (UMR 7334), Institut Mat\'{e}riaux Micro\'{e}lectronique et Nanosciences de Provence,  Marseille, France.}

\newcommand{\affLASIR}{CNRS, Universit\'{e} de Lille, LASIRE (UMR 8516), Laboratoire de Spectrochimie Infrarouge Raman et Environnement, Villeneuve d'Ascq, France}
\newcommand{\affISCR}{Universit\'{e} de Rennes, CNRS, ISCR UMR 6226, F-35042 Rennes, France.}
\newcommand{\affISM}{CNRS, Aix-Marseille Universit\'{e},  Centrale Marseille, ISM2, Institut des science mol\'{e}culaire de marseille,  Marseille, France.}

\newcommand{\1}{\textcolor{black}} 
\begin{document}
	
\title{Electron spins interaction in the spin-Peierls phase of the organic spin chain \DMTTFX ($X$ = Cl, Br, I) }

\author{L.~Soriano}\affiliation{\affIMNP}
\author{O.~Pilone}\affiliation{\affIMNP}
\author{M.D.~Kuz'min}\affiliation{\affIMNP}
\author{H.~Vezin}\affiliation{\affLASIR}
\author{O.~Jeannin}\affiliation{\affISCR}
\author{M.~Fourmigu\'{e}}\affiliation{\affISCR}
\author{M.~Orio}\affiliation{\affISM}
\author{S.~Bertaina}\email{sylvain.bertaina@cnrs.fr}\affiliation{\affIMNP}
\date{\today}

\begin{abstract}
We investigate the electron spin resonance of the organic spin-Peierls chain
\DMTTFX with $X$ = Cl, Br and I. We describe the temperature dependence of the
spin gap during the phase transition and quantify the dimerization parameter
$\delta$. At the lowest temperatures, the susceptibility is governed by defects
in the spin dimerized chain. Such strongly correlated defects are the
consequence of breaks in the translational symmetry of the chain. In the
vicinity of the defects the spins are polarized antiferomagnetically forming a
magnetic soliton: a spin $\frac{1}{2}$ quasi-particle of size ruled by $\delta$
pinned to the defects. For \Br and \Cl, we show that the one-half of the total number of solitons
are in isolation (as singles) whereas the other half form pairs (soliton
dimers)with a strong magnetic coupling. The Rabi oscillations of both the
single-soliton and the soliton-dimer are observed, which is a prerequisite in
the context of quantum information.
\end{abstract}
\maketitle
\section{Introduction}

The physics of spin $S=\frac{1}{2}$ chains remains extremely rich because their low
dimension leads to pronounced influence of the electronic correlation and allows
the interplay between magnetic, electronic and lattice degree of freedom
\cite{White1983,Vasiliev2019}. In particular, in $S=\frac{1}{2}$ antiferromagnetic
Heisenberg spin chains, the quantum fluctuation prevents long-range order and the
ground state is gapless \cite{Bethe1931}. However, this sate is unstable and a
weak coupling with the other chains or with the lattice opens a gap in the
magnetic spectrum and leads to a long-range order (antiferromagnetic order) or to
dimerization (spin-Peierls). The effect of defects in 1D spin systems continue
to be actively studied because the break in the translation symmetry deeply
alters the magnetic properties of the host materials \cite{Alloul2009}. In the
spin-Peierls infinite chains, the ground state is a singlet ($S=0$) separated from the 
quasi continuum by a gap \cite{Foury-Leylekian2009,Regnault1996}. The break in
the translational symmetry, like a chain-end or a stacking fault, alters the spins
in the vicinity creating a magnetic soliton (spin-$\frac{1}{2}$ quasiparticle
made of many correlated spins) pinned to the defects
\cite{Khomskii1997,Nishino2000a}. As a consequence of this many-body
spin-$\frac{1}{2}$ soliton formation, the ground state is a doublet separated from the quasicontinnum by a gap \cite{Sorensen1998}. Such a structure is of particular
interest because its energy levels are comparable to single molecular magnet SMM
like V15 \cite{Shim2012,Tsukerblat2006,Soriano2020} with unconventional quantum
coherence properties \cite{Bertaina2014a,Zeisner2019a} which make it an
interesting potential qubit \cite{Bose2003, CamposVenuti2006}.

Organic one-dimensional conductors were extensively studied over the past
decades due to the richness of the phase diagram. One of the most famous is the
Fabre salt (TMTTF)$_2X$, where $X$ is a counter anion. Depending on the
temperature, the pressure and the nature of $X$, (TMTTF)$_2X$ can be
a metallic/insulator/superconductor uniform/dimerized spin
chain, N\'{e}el-/charge-/anion- ordered (\cite{Dressel2007,Coulon2004,Pouget2012,
	deSouza2013, Pouget2016, Pouget2018}). The defects in the spin chains have been
observed by electron spin resonance (ESR) \cite{Bertaina2014,Dutoit2018} but due to a low spin-Peierls
transition temperature \TSP and to a high homogeneity of the ESR line, the
quantum coherence study is limited.

The \DMTTFX family compounds, with $X$ = Cl, Br and I, have \TSP$\sim 50$~K higher than \PF (\TSP= 19~K) which itself has the highest \TSP of the (TMTTF)$_2X$ series. Moreover electron spin echo of the pined soliton has been reported
\cite{Zeisner2019a}. \DMTTFX was first synthesized many decades ago
\cite{Abderraba1983} but intensive studies have been published recently
\cite{Fourmigue2008} with the construction of the phase diagram  
\cite{Foury-Leylekian2011,Auban-Senzier2012}, the solid solution with different
counter anion \cite{Reinheimer2012} and the ESR study\cite{Zeisner2019a,Soriano2020}

\begin{figure} \centering
	\includegraphics[width=\mycol\linewidth]{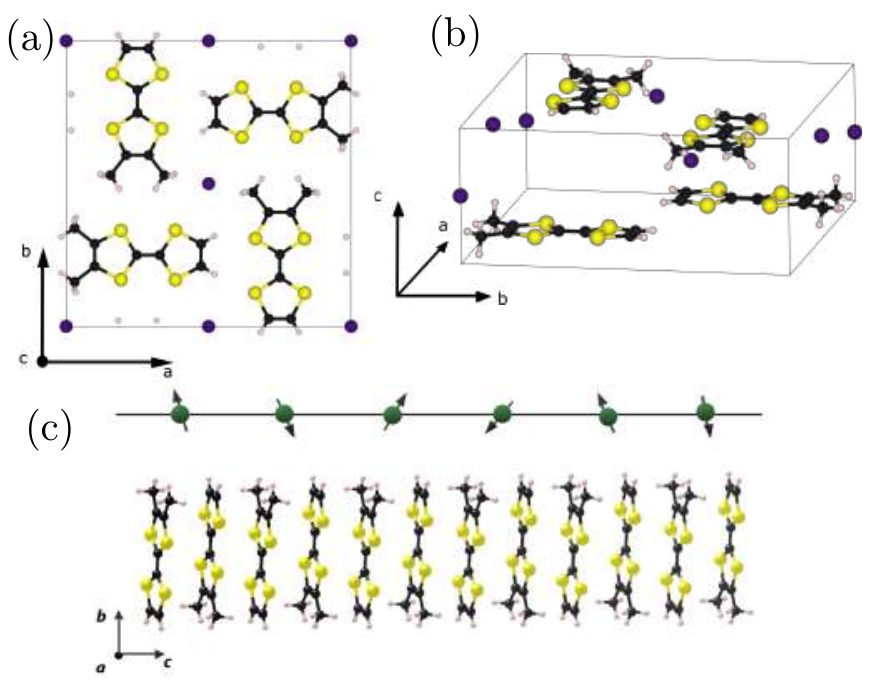}
	\caption{Crystallographic structure of \DMTTFX. (a) View of the $ab$ plane and
		(b) perspective view of the unit cell with four DMTTF in equivalent positions
		rotated by 90$^\circ$. The DMTTF molecules are stacked along $c$ axis but only 4
		molecules are shown to avoid confusions. (c) Stacks of (DMTTF)$_2$ along the $c$
		axis. Each pair of (DMTTF)$_2$ shares a spin $S=\frac{1}{2}$ forming a spin
		chain along $c$.      } \label{fig:Structure} \end{figure}

The three systems \DMTTFX crystallize in the same space group $I\bar{4}2d$ (no.
122) with cell parameters  $a=b=$ 16.93~\AA, 17.09~\AA, 17.40~\AA\xspace and
$c=$ 7.040~\AA, 7.058~\AA, 7.098~\AA\xspace  for \Cl, \Br and \I respectively. The
halide anions $X$ are in position $\bar{4}$ while $o-$DMTTF molecules lie on the
two-fold axis forming a stack in the direction $c$. Each linear stack is turned
by 90$^\circ$ with respect to its neighbors, as shown in Fig. \ref{fig:Structure}. This ``chessboard''
structure has a consequence of a very weak inter-stack interactions compared to
the parallel stack of the famous (TMTTF)$_2X$, confirmed by the highly anisotropic
conductivity of \DMTTFX \cite{Auban-Senzier2012,Foury-Leylekian2011}. Each pair
of  $o-$(DMTTF)$_2$ shares a spin $S=\frac{1}{2}$ and forms a quasi-isotropic
Heisenberg spin chain along $c$-axis. When $T<T_{SP}$ the displacement of the pairs
of $o-$(DMTTF)$_2$  creates the tetramerization of the structure along the chain
axis. To describe the systems in the spin-Peierls phase we will use the
$S=\frac{1}{2}$ alternating-exchange Heisenberg chain Hamiltonian.
\begin{equation}
	\label{eq:HAFM-Hamiltonian}
	\mathcal{H} = \sum_{i}\left[ J(1+\delta)\boldsymbol{S}_{2i-1}\cdot \boldsymbol{S}_{2i} + J(1-\delta)\boldsymbol{S}_{2i}\cdot\boldsymbol{S}_{2i+1}\right]  \ \ \ .
\end{equation}
Here $J>0$ is the AFM isotropic Heisenberg exchange integral and $\delta$ the explicit
alternation parameter. In a spin-Peierls system, $\delta$ is related to the
elastic energy of the lattice and the magneto-elastic coupling, which are both 
responsible for the dimerization of the chain. The effects of impurities in
quantum spin chains have been actively studied in the past. Thanks to the
improvement of numerical methods, studies using exact diagonalization
\cite{Hansen1999}, Quantum Monte Carlo \cite{Eggert1992,Nishino2000} and Density
Matrix Renormalization Group \cite{Augier1999} exhibit the many-body nature of
non-magnetic defects.

\begin{figure} \centering \includegraphics[width=\mycol\linewidth]{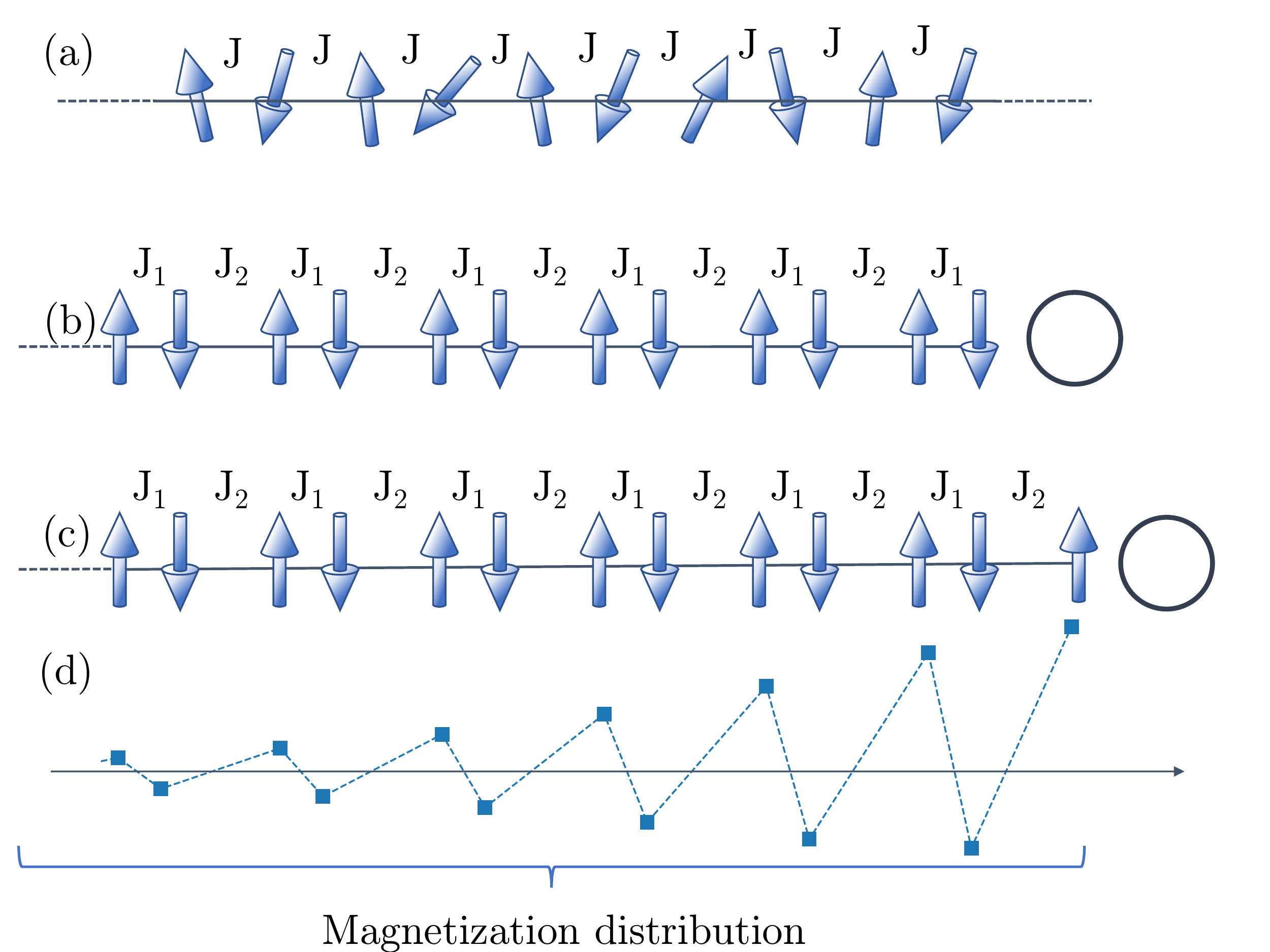}
	\caption{Schematic representation of the spin chain during the spin-Peierls
		transition and the defect-induced soliton formation. (a) For $T>T_{SP}$, the
		uniform Heisenberg spin chain is made of constant exchange coupling $J$ between
		equidistant nearest-neighbor spins. (b)and (c) For  $T<T_{SP}$ the bond lengths
		within the spin chain is modulated being alternatively shorter or longer leading
		to an exchange coupling stronger ($J_1=J(1+\delta)$) or weaker
		($J_1=J(1-\delta)$) respectively. The presence of non-magnetic defects (empty
		circle) create finite chains. (b) If the chain contains an even number of spin,
		they form pairs and the ground state is non-magnetic but (c) if it contains an
		odd number of spins, one spin remained unpaired and the ground state is
		magnetic. (d) Local magnetization of the chin in (c) calculated by DMRG
		\cite{Bauer2011} with $\delta$=0.1. } \label{fig:Model1} \end{figure}

Fig. \ref{fig:Model1} summarizes the spin-Peierls transition and the effect of
non-magnetic impurities. In the case of a finite length dimerized chain, 
parity plays an important role in determining the magnetic properties. In case of an even
number of spins (Fig.\ref{fig:Model1}(b)), they create pairs of spins and the
ground state is non-magnetic with a large gap comparable to that of the infinite
spin-Peierls chain. In case of an odd number of spins (Fig.\ref{fig:Model1}(c)),
one spin remains unpaired and the ground state is magnetic, separated from the next
states by a gap comparable to that of the infinite chain. However, contrary to a
magnetic impurity inside a non-magnetic medium \cite{Wolfowicz2021}, the unpaired spin is
correlated to the rest of the chain and the local polarization is spread over
many neighbor spins, forming a magnetic soliton (as calculated by DMRG \cite{Bauer2011}
with $\delta$=0.1 Fig.\ref{fig:Model1}(d)).


Since the ground state of the spin chain defects is a doublet of effective spin
$S=\frac{1}{2}$ it should be quantitatively accessible through temperature
dependent static susceptibility by means of SQUID magnetometry or electron spin
resonance (ESR). The former method suffers from the impossibility to separate the effect of spin chain defects from other
extrinsic contribution (dirt, paramagnetic impurities...). ESR, by adding the spectral dimension, can separate the different
contributions.

In this paper we present an ESR investigation of \DMTTFX with $X$=Cl, Br and I
from room temperature down to $T=5$~K. \1{ESR measurements on these systems have been reported \cite{Zeisner2019a,Soriano2020}, but with a weak density of data points and, as we will explain in the following, an incorrect analysis of the gap.} First, in Sec. \ref{sec:Susc} we present
the magnetic susceptibility and extract a quantitative estimation of the
microscopic parameters, such as the temperature dependence of the intra-chain
coupling $J_{eff}(T)$ (Sec. \ref{sec:HT}), the temperature dependence of the
dimerization parameter and the gap across the spin-Peierls transition(Sec.
\ref{sec:Dim}), and the content of spin chain defects. Then, in Sec.
\ref{sec:ElecInter} by means of a continuous wave (CW) and pulsed ESR we study the
dynamics of isolated and coupled magnetic solitons pinned by non-magnetic
defects of the chain.

\section{Experimental Details} Single crystals of \Br, \I and \Cl have been
grown by electro-crystallization using the standard procedure described in Ref.
[\onlinecite{Fourmigue2008}]. The crystals are needle shape with the chain axis
along the long length of the needle. The largest crystals have been used in the
low temperature ESR study where the signal of the defects are observed but weak
due to the low concentration of strongly correlated defects. The typical size
was 0.2$\times$0.2$\times$3 mm$^3$ along the $a$, $b$ and $c$ axes. To avoid
effects of temperature cycle history, a fresh sample was used for each series of
measurements. The samples were glued on suprasil quartz rode using a small
amount of Apiezon grease on one side of the samples to avoid too much stress
while sweeping the temperature.

CW-ESR measurements were performed using a conventional Bruker EMX spectrometer
operating in X-band with microwave frequencies of about  $f_{mw}=9.387$~GHz.
This spectrometer is equipped with a He-flow cryostat (ESR900) and a cryogen-free cooler (Bruker Stinger
)which operates down to 7~K. The angular dependence of ESR with respect to the static
field  was measured using an automatic goniometer installed on the spectrometer.
The angle $\theta=0^\circ$ corresponds to $H\| c$. We paid particular
attention to the low temperature regime where the relaxation could be long.
Therefore, the microwave power was set low ($< 1$~mW) to prevent the ESR signal
from saturation. The field modulation was set under 1 G to avoid the distortion of
ESR lines due to over-modulation effect.

For pulsed ESR experiments, we used a Bruker Elexsys E580 spectrometer equipped
with a cryogen-free cryostat. The Rabi oscillation measurements were performed
with the external static field $H_0$  applied along the $c$-axis and the
microwave frequency of $f_{mw}=9.693$~GHz.The amplitude of the microwave field
$h_{mw}$ was calibrated using a $S=\frac{1}{2}$ radical. The sequence used was
the following: $p(t)-T-\pi/2-\tau-\pi$-echo, with $p(t)$ the Rabi pulse of a
duration $t$, $T\gg T_2$ the waiting time and $\pi/2-\tau-\pi$-echo is the standard Hahn
echo used to probe $\left\langle S_z \right\rangle(t)$. A $S=\frac{1}{2}$ radical (DPPH) is used to calibrate the amplitude of the microwave field.

\section{Results and Discussion}

\subsection{Susceptibility}\label{sec:Susc}

Let us first describe the magnetic susceptibility extracted from ESR
measurements. In the linear response theory, the susceptibility from ESR
$\chi_s$ is usually related to the spectral intensity using the Kramers
Kr\"{o}nig relation: $\chi_s=\int \chi"(\omega) d\omega$ which in case of small
linewidth and anisotropy becomes $\chi_s=\int I_{ESR}(H) dH$ . This is true when
the ESR signal is due to absorption only, but here, \DMTTFX is a conductor at high
temperature and the dispersion induced by the conductivity has to be taken into
account \cite{Coulon2007}. We use eq.\eqref{eq:fit} to fit our ESR data:

\begin{equation}\label{eq:fit} 
	I_{ESR}= A\left(\frac{\Gamma\cos \phi }{\Gamma^2+(H-H_0)^2} + \frac{(H-H_0)\sin \phi}{\Gamma^2 +(H-H_0)^2} \right)
\end{equation} 
where $A$ is the amplitude of the signal and is directly
proportional to the magnetic susceptibility, $\Gamma$ is half-width at half-maximum,
 $H_0$ is the resonance field and $\phi$ the angle of dispersion. This
fit procedure is very accurate for all orientations and for T$>$20~K. However
below 20~K the signal attributed to the spin chain defects cannot be fitted by eq.
\eqref{eq:fit} with a good accuracy and we decided to use the standard double
integration of the signal to obtain $\chi_s$. This is possible since at these
temperatures the \DMTTFX family is an insulator \cite{Foury-Leylekian2011}.

\begin{figure} \centering	\includegraphics[width=\mycol\linewidth]{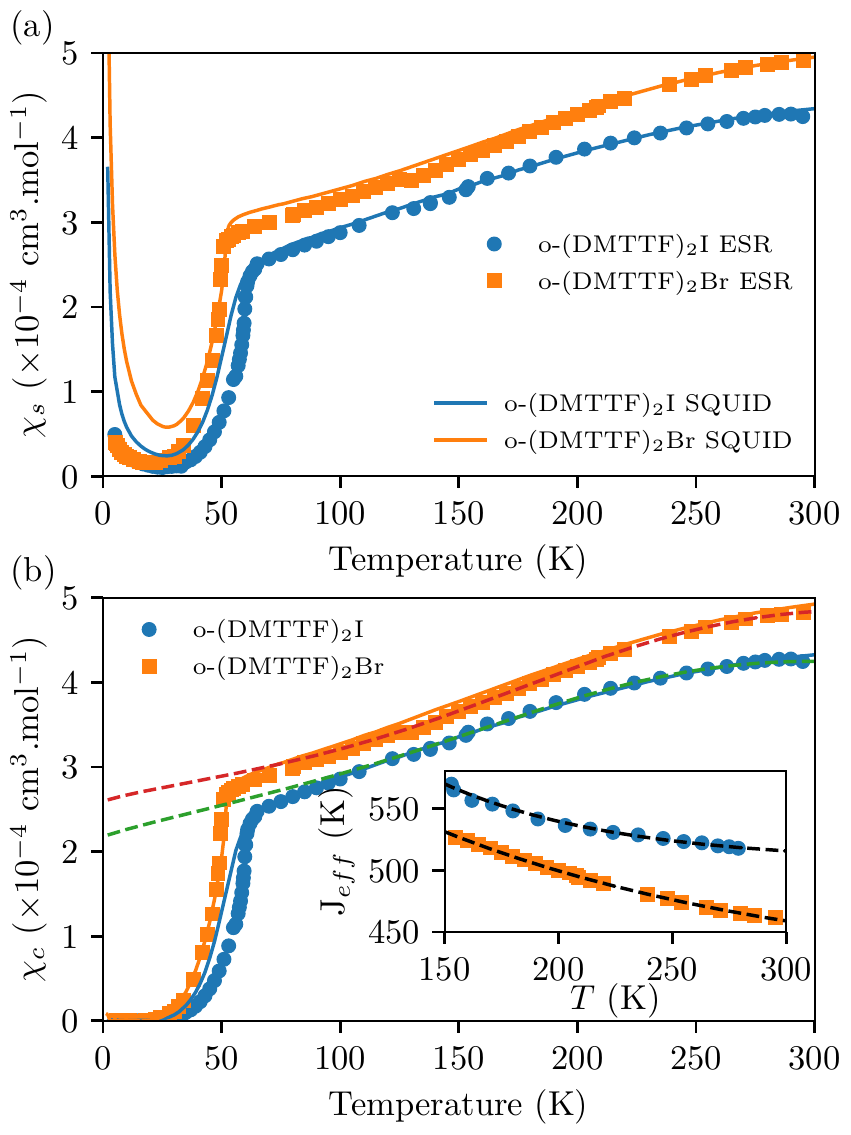} \caption{(a)
		Temperature dependence of the spin susceptibility $\chi_s$ deduced from ESR in
		\Br (squares) and \I (circles). The plain lines are SQUID data adapted from ref
		[\onlinecite{Foury-Leylekian2011}]. (b) Temperature dependence of the spin
		susceptibility corrected for the Curie tail of $S=1/2$ defects $\chi_c$ (see
		table \ref{tab:impurity}). The dashed lines are the theoretical values of the
		susceptibility using exchange constants $J_{eff}$ presented in the inset and
		extrapolated to low temperature.   } \label{fig:Chi_HT} \end{figure}

Fig.\ref{fig:Chi_HT} (a) shows the temperature dependence of the susceptibility
extracted from the ESR measurements $\chi_{ESR}(T)$ of \I and \Br (\1{\Cl is very close to \Br and is reported in the \textit{Supplementary Materials} \footnote[1]{See Supplemental Material online for ESR spectra examples, complementary susceptibility data, Rabi field sweep for the 3 compounds, residue analysis of X=Cl and X=I and python code used for the Johnston \etal model.} to avoid overloading of Fig.\ref{fig:Chi_HT}} ).  $\chi_{ESR}$ is usually in arbitrary units since it depends on the
experimental parameters. However, using reference data, it is possible to renormalize
$\chi_{ESR}$ to absolute units. In our case, we renormalized $\chi_{ESR}$
using independent SQUID measurements of the same compounds
\cite{Foury-Leylekian2011}. In Fig. \ref{fig:Chi_HT}, the squares and circles
are ESR data while the lines are SQUID measurements from
Ref.\cite{Foury-Leylekian2011}. Above \TSP the susceptibilities measured by ESR and by 
the SQUID are very similar. However,  at low temperature the Curie tails induces by magnetic impurities are clearly different.

\begin{table}[h] \caption{Concentration of $S=1/2$ defects extracted from the
		Curie like behavior at low temperature assuming g=2\label{tab:impurity}}
	\begin{center} \begin{tabular}{lcc} \hline
			\noalign{\smallskip}\hline\noalign{\smallskip} & SQUID $10^{-4}.at^{-1}$
			\cite{Foury-Leylekian2011} &ESR $10^{-4}.at^{-1}$  \\
			\noalign{\smallskip}\hline\noalign{\smallskip} \Cl & 25 & 5.6 \\ \Br& 37 & 6.6\\
			\I & 14 & 4\\ \noalign{\smallskip}\hline\hline \end{tabular} \end{center}
\end{table}

Table \ref{tab:impurity} shows the concentration of defects/impurities extracted
from the Curie behavior at low temperature assuming $S=1/2$ and g=2.
Clearly, the quantity of impurities is larger in the SQUID measurements than in
ESR. This is not surprising since SQUID measurements are not selective and yield the
total magnetic moment of a bulk  sample, while ESR is highly selective and provides
information on a particular kind of impurities.
Fig. \ref{fig:Chi_HT}(b) shows the spin chain susctibility $\chi_c$ upon the deduction of the low-temperature Curie-tails.
 For the full range of temperature and in the
limit of experimental error,  the SQUID and ESR data on \Br are identical.
This is less clear for \I. For $T>$\TSP SQUID and ESR susceptibility are
identical, however, below \TSP, $\chi_c$ from ESR shows a much stronger
temperature dependence than the SQUID one. We think that this difference is due
to the nature of \I which should be close to a highly pressure-sensitive region of the phase diagram. As noted in Ref.
\cite{Foury-Leylekian2011} \I is difficult to place on the phase diagram. The
authors observed a charge density wave (CDW) transition at $T_C=47$~K and no gap
in the ESR indicating a position in the high-pressure zone of the phase diagram,
while we observe a clear spin-Peierls transition at \TSP=63~K and a behavior
comparable to that of \Br and \Cl suggesting a pressure lower than expected. This
difference might be due to the method of gluing the samples, which  can induce
different strains at low temperature.

In the following we describe the susceptibility in both the uniform spin chain
phase ($T>$\TSP) and the dimerized phase ($T<$\TSP) using a method developed by
Johnston\etal \cite{Johnston2000}. My means of Quantum Monte Carlo (QMC) and
transfer-matrix density-matrix renormalization group (TMRG) they unified and
improved the theoretical predictions developed  for the dimerized spin chain  by Bulaevskii
\cite{Bulaevskii1969} and for the uniform spin chain by Bonner and Fisher
\cite{Bonner1964}, Eggert Affleck and Takahashi \cite{Eggert1994a}, and Kl\"{u}mper
and Johnston \cite{Klumper2000} .

\subsubsection{Uniform chain susceptibility}\label{sec:HT}

In the uniform spin chain regime ( $T>T_{SP}$ ), all the models cited above failed
to describe the susceptibility of \DMTTFX. This anomaly has been observed in
(TMTTF)$_2$PF$_6$ \cite{Salameh2011}, deuterated
(TMTTF)$_2$PF$_6$D$_{12}$\cite{Pouget2017}, 
(TMTTF)$_2$SbF$_6$\cite{Foury-Leylekian2009},(TMTTF)$_2$AsF$_6$ \cite{Dumm2004} 
and has been attributed to thermal expansion. All the theoretical models describe susceptibility at constant volume
($[\chi]_V$) while measurements are performed at constant pressure ($[\chi]_P$).
A method to convert the temperature dependence of ($[\chi]_P$) to ($[\chi]_V$)
was developed by Wzietek\etal\cite{Wzietek1993} in the case of (TMTSF)$_2$PF$_6$
by performing X-ray and nuclear magnetic resonance (NMR) under pressure. This
method is laborious and suffers from the arbitrarity of choice of reference temperature volume. We choose a different approach by extracting
the exchange constant $J$ as function of the temperature at constant pressure.
This is possible with the Johnston\etal method but needs very accurate absolute
measurements.

To describe the uniform spin chain phase we use the following method: (a) we
choose a range of temperature far enough from \TSP to avoid fluctuations of
spin-Peierls and ensure that the alternating parameter $\delta$ vanishes, in our
case T=150~K to 300~K, which is at least 3 times \TSP. (b) We use the Pad\'{e}
approximant and the coefficients provided in Table I of Ref.\cite{Johnston2000}
to extract the exchange constant $J_{eff}(T)$ for each temperature in this
range. The result is provided in the inset of Fig.\ref{fig:Chi_HT}(b) for \Br
and \I. We observe that the effective exchange coupling $J_{eff}$ decreases as T increases.
Such behavior has been noticed in NaV$_2$O$_5$
\cite{Johnston2000} and qualitatively explained by Sandvik\etal
\cite{Sandvik1997} and K{\"u}hne \etal \cite{Kuhne1999} by including
dispersionless phonons (Einstein phonons) linearly coupled to the spin chain. In
our case, a more direct effect is the variation of cell volume observed by
changing the temperature. Radical organic salts are known to have a large
thermal expansion compared to inorganic metal oxides. In (TMTYF)$_2$XF$_6$ (with
X=Sb, As, P and Y=S or Se)\cite{Granier1988,Furukawa2009,deSouza2008} the
variation of the chain length is about 3\%. For \Br and \I, the X-ray
diffraction measurements show a linear temperature dependence of the cell
parameters, theincrase of $c$ change between $T=100$~K and room
temperature of about 2\% (see Appendix \ref{sec:A_XRAY} for details).

To estimate the effect of the variation of the cell parameter $c$, we used the
molecular DFT calculation on a minimal dimer cell (see Appendix \ref{sec:A_DFT}).
A 2\% change of the distance between 2 $o$-(DMTTF)$_2$ molecules leads to
13\% change of the exchange constant, in agreement with the values of
$J_{eff}$ presented in the inset of Fig.\ref{fig:Chi_HT}(b)

By incorporating the effective exchange coupling constant $J_{eff}$ calculated
from $T=150$~K to $T=300$~K and extrapolated to lower temperature into the
uniform spin chain model  \cite{Johnston2001} we obtain the colored dashed lines
in Fig. \ref{fig:Chi_HT}(b). We notice that susceptibility calculated using the
extrapolated  $J_{eff}$ reproduces the experimental data with a good accuracy
down to $T=100$~K. A small discrepancy observed for $T<100$~K can be attributed to the
opening of a pseudogap, which is in agreement with the temperature where weak
diffuse X-ray  scattering lines were observed \cite{Foury-Leylekian2011}- a sign of a pretransitional effect.

\subsubsection{Dimerization of the spin chain }\label{sec:Dim}

For $T<$\TSP the gapless Heisenberg uniform spin chain progressively enters in a
non-magnetic gapped stated ($S=0$) via a spin-Peierls transition
\cite{Foury-Leylekian2011}. The susceptibility extracted from our ESR study
($\chi_{ESR}$) below \TSP shows no significant difference from the DC
susceptibility from previous SQUID measurements for either \Br (Fig. \ref{fig:Chi_HT})
or \Cl (see SI). However, for \I  shows a temperature dependence
comparable to that in \Br and \Cl (with a higher \TSP) but is significantly different
from the SQUID data.

\begin{figure} \centering \includegraphics[width=\mycol\linewidth]{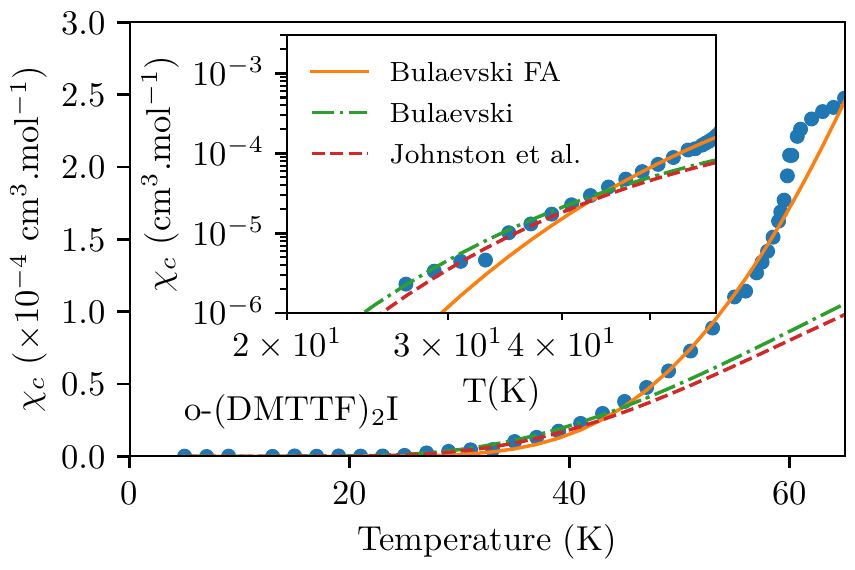}
	\caption{Temperature dependence of the spin susceptibility in the
		low-temperature region for \I. Below \TSP=63~K the susceptibility drops to the
		non-magnetic spin-Peierls state. The plain orange line is the best fit using eq.
		\eqref{eq:Bulaeski} taking $\alpha$ and $\Delta$ as two independent fit
		parameters. The green dashed dot and red dashed lines are the best fits to the
		Bulaevski and Johnston \etal model respectively using only $\delta$ as a fit
		parameter. The inset is the log-log scale of the figure, magnifying the
		discrepancy of the Bulaeveski FA model at low temperature. Analysis for \Br and \Cl are provided in \cite{Note1}}
	\label{fig:ChiLT} \end{figure}

The determination of the spin gap $\Delta$ is of fundamental importance since it is directly
related to the dimerization parameter $\delta$ \cite{Barnes1999,Augier1997}.
However it is a tricky problem since it depends on the microscopic model used. 
The model that is the most widely used in the literature was developed by Bulaevski
\cite{Bulaevskii1969} and consists in the analytical calculation of the
susceptibility in the Hartree-Fock approximation. He calculated the magnon
dispersion of coupled dimers and in the low temperature approximation he
provided a simple two-parameter form of the susceptibility:

\begin{equation}\label{eq:Bulaeski}
	\chi_s(T)=\frac{\mathcal{N}_ag^2\mu_B^2}{k_BT}.\alpha(\delta)\exp\left(-\frac{\Delta_B(\delta)J_1}{k_B T}\right) \end{equation}

With $J_1=J(1+\delta)$, in cgs units and with $g\approx2$ the pre-factor is
close to  0.375. \1{$\alpha$ is the amplitude factor and $\Delta_B.J_1$ is the gap.}Despite its simplicity, this model shows remarkably good
agreement with modern numerical approach for large dimerization
($\delta>0.5$)\cite{Barnes1999,Johnston2000} but the agreement becomes
progressively worse as $\delta$ decreases and break down for $\delta<0.1$. In
the latter case, the reason is that the magnon dispersion minimum is not at $k=0$
\cite{Barnes1999}. Independently of the range of validity of $\delta$ in the
Bulaevski's model, one should use eq. \eqref{eq:Bulaeski} with caution.
First of all, Bulaevski's approach is essentially a low-temperature one; the dimerization parameter $\delta$ and the energy gap $\Delta$ are independent of temperature. This is, of course, not fulfilled in the entire range of existence of the spin-Peierls state. 
Orignac \etal \cite{Orignac2004} showed
that $\Delta(T)=\Delta(0)$ for $T<0.5$\TSP. The second point to take care of is that
the two parameters $\alpha$ (the amplitude) and $\Delta$ (the spin gap) are not independent. Rather, both are functions of $\delta$, tabulated in Ref. \cite{Bulaevskii1969,Note1}. The latter
fact has been often neglected, leading to an incorrect use of the Bulaevski model
\cite{Hase1993, Salameh2011, Zeisner2019a}.

An example of determination of the dimerization parameter $\delta$ by different
methods using $\chi_{ESR}$ for \I is given in Fig. \ref{fig:ChiLT}.
The model labeled "Bulaevski free amplitude (Bul. FA)" corresponds to eq.
\eqref{eq:Bulaeski} with \1{$\alpha(\delta)$ replaced by $\alpha$, which is now a free parameter independent of $\delta$, while $\Delta_B(\delta)$ remains a function of $\delta$. It is clear that this model is incorrect and should not be used but it was applied in the past to extract the dimerization parameter of spin-Peierls systems\cite{Dumm2000a,Salameh2011,Soriano2020} and had certainly provided an overestimated $\delta$ as we will se in the following.} The fit labeled  "Bulaevski" is
eq. \eqref{eq:Bulaeski}  with only  $\delta$ as a free parameter as it should be used. Finally, "Johnston \etal"
is a direct numerical calculation of the susceptibility using TMRG with only
$\delta$ as a free parameter. \1{"Johnston \etal" use the Pad\'{e}
	approximant and the coefficients provided in Table I of Ref.\cite{Johnston2000} and does not suffer of the approximation made by Bulaevski for his analytical description \eqref{eq:Bulaeski}} At first sight Bul. FA seems a better fit, but a
closer look at low temperature on the log-log scale (Fig. \ref{fig:ChiLT} inset)
detects an important discrepancy with the data. On the contrary, the two other
models used, correctly show a very good agreement with experimental data for
$T<40$~K. At higher temperature the dimerization $\delta(T)$ decreases and the
models cannot be used in the current form any more.

\begin{table}[h] \caption{Dimerimzation parameter $\delta (T=0)$ extracted from
		three different models: Bulaevski free amplitude (Bul. FA) correspond to eq.
		\eqref{eq:Bulaeski} with $\alpha$ left as a free parameter, Bulaevski model is
		\eqref{eq:Bulaeski} with $\delta$ is the only fit parameter and the Johnston
		\etal model. (see text for details)    \label{tab:Gap}} \begin{center}
		\begin{tabular}{lcccc} \hline \noalign{\smallskip}\hline\noalign{\smallskip} &
			Bul. FA  &Bulaevski \cite{Bulaevskii1969} & Johnston \etal \cite{Johnston2000} 
			\\ \noalign{\smallskip}\hline\noalign{\smallskip} \Cl & 0.17 & 0.088  & 0.083\\
			\Br& 0.14 & 0.085 &0.080\\ \I & 0.18 & 0.10 &0.096\\
			\noalign{\smallskip}\hline\hline \end{tabular} \end{center} \end{table}

Let us note a large overestimation of $\delta$ in the Bul. FA fit while
Bulaevski and Johnston \etal fits produce rather consistent values of $\delta$, those of Bulaevski being slightly higher.

The relation between the magnetic gap $\Delta$ and the dimerization parameter
$\delta$ is a central problem in understanding the microscopic properties of
dimerized spin chains. It has been intensively studied in the past, following the
development of modern analytical (like bosonization) or numerical (DMRG)
approaches. It was shown \cite{Black1981} that the critical behavior
$\Delta\sim\delta^{2/3}$ \cite{Cross1979} must be corrected to
$\Delta\sim\delta^{2/3}/|\ln \delta|^{1/2}$  \cite{Spronken1986}. Note the
absence of prefactors in the early developments in the field. More recently
numerical developments have led to quantitative description of the spin gap,
like $\Delta/J=2\delta^{3/4}$ by Barnes \etal \cite{Barnes1999},
$\Delta/J=1.94\delta^{0.73}$ by Papenbrock \etal \cite{Papenbrock2003}, and the
elegant analytical solution proposed by Orignac \cite{Orignac2004a}
$\Delta/J=1.723\delta^{2/3}$. The validity of all these formulas depends on the
range of $\delta$, for $\delta\sim 0.1$ the agreement lies within 5\%. \\

The particularity of spin-Peierls systems is to have a temperature-dependent
gap. Far below the transition temperature the gap is independent of temperature
and the method presented above is sufficient to estimate $\Delta$. However, for
$0.5T_{SP}\lesssim T<T_{SP}$ the gap has to be treated more carefully. Here we treat the
temperature dependence of the spin gap $\Delta(T)$ following Johnston
\etal \cite{Johnston2000}. 
Proceeding from the temperature dependence of the exchange constant found in \ref{sec:HT}, see the inset of Fig. \ref{fig:Chi_HT}(b), we extrapolate $J_{eff}(T)$ towards lower temperatures and evaluate it just above \TSP. Let us denote this value by $J^*_{SP}$). We obtain $J^*_{SP}\simeq
600$~K for \Cl and \Br and $J^*_{SP}\simeq 670$~K for \I. Then $\delta(T)$ is
computed using the fit function for the alternating-exchange chain (see Table II
of Ref. \onlinecite{Johnston2000}) by finding the root for $\delta$ at each
experimental point. Finally, the temperature dependence of the spin gap
$\Delta$(T) is computed using the Barnes \etal \cite{Barnes1999} relation,
$\Delta(T)=2\delta^{3/4} J^*_{SP}$.

\begin{figure} \centering
	\includegraphics[width=\mycol\linewidth]{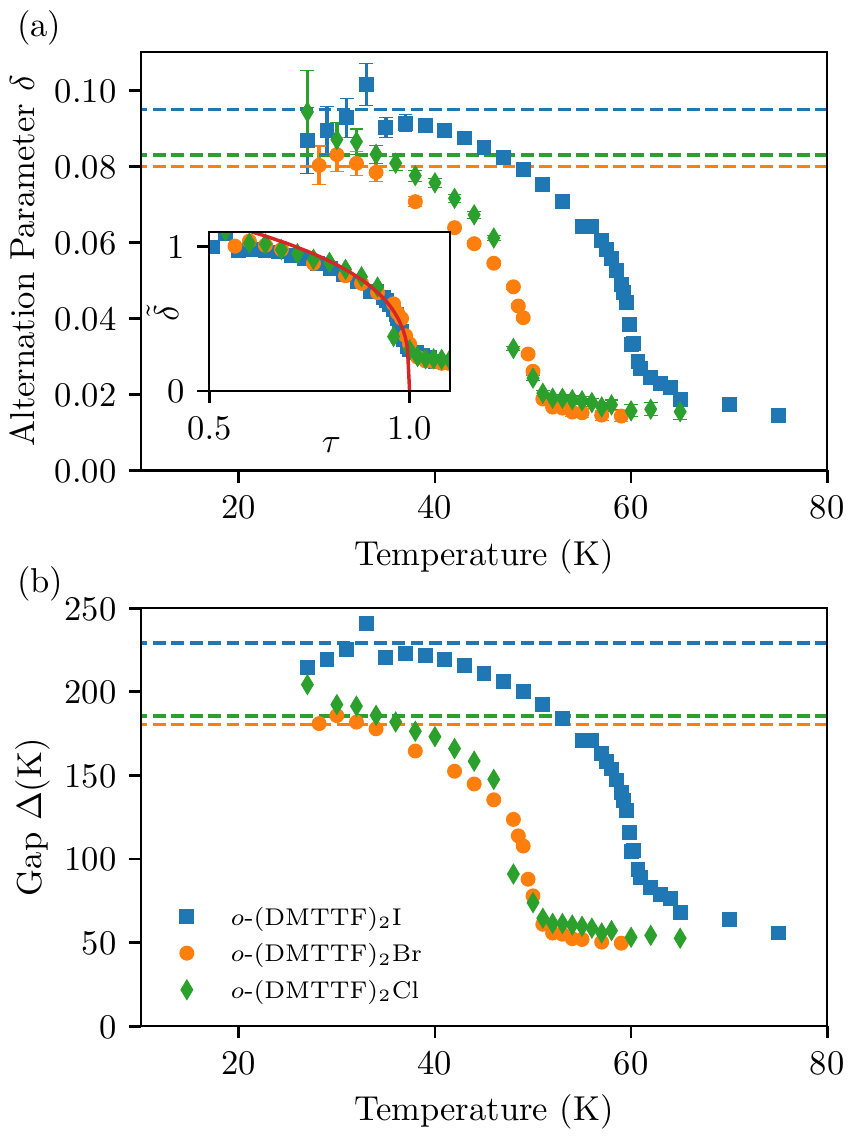} \caption{Temperature
		dependence of (a) the dimerization parameter $\delta$ and (b) the spin gap
		$\Delta$ for \I (blue squares), \Br (orange circles) and \Cl (green diamonds). The
		dashed lines represent the values estimated at $T=0$~K (see Table
		\ref{tab:Gap}). The inset is the reduced temperature dependence
		($\tau=T/T_{SP}$) of the reduced dimerization parameter
		$\tilde{\delta}=\delta(T)/\delta(0)$.} \label{fig:AlterGap} \end{figure}

The temperature dependence of the alternation parameter $\delta(T)$ and the
corresponding spin gap $\Delta(T)$ for \I, \Br and \Cl is given in Fig.
\ref{fig:AlterGap}. \1{The method used to extract $\delta(T)$ is explained in Ref.\cite{Johnston2000} and summarized here: for each value of the susceptibility $\chi_c(T)$ we solve the equation $\chi_J(\delta,J,T)=\chi_c(T)$ for non-vanishing value of $\chi_c$. Where $\chi_J(\delta,J,T)$ is the susceptibility provided by the Johnston \etal model. Knowing $J$ from the Sec. \ref{sec:HT} we obtain $\delta$ as function of temperature.   }
The dashed lines correspond to the values of $\delta$ and
$\Delta$ at $T=0$~K taken from Fig. \ref{fig:ChiLT} and Tab. \ref{tab:Gap}.
We notice that for \1{ $T<0.75 T_{SP}$ }the dashed lines are the asymptotes of
$\delta(T)$ and $\Delta(T)$. It is worthwhile noticing that the method described
above is valid when the susceptibility is non zero. This is the reason why no value can be provided below
about 25~K. \1{ We show that below 0.75\TSP}, $\delta$ and
$\Delta$ are independent of temperature and can be estimated by their values at $T=0$~K
values. Above 0.75\TSP the dimerization parameter $\delta(T)$ and the spin gap
$\Delta(T)$ decrease as $T$ approaches \TSP, but they do not vanish at $T=$\TSP. The
data in Fig. \ref{fig:AlterGap} clearly show the existence of spin dimerization
fluctuations and a spin pseudogap above \TSP for the three \DMTTFX compounds of about 20\% and 30\% of $\delta(0)$ and
$\Delta(0)$, respectively. The fluctuation effects above \TSP seem to persist at
high temperature and show a pretransitional effect of the lattice confirming
the observation reported by X-ray diffusion scattering
\cite{Foury-Leylekian2011}. Precursor effects above \TSP have been reported in
both organic \cite{Rovira1995} and inorganic \cite{Fertey1998,Smirnov1998a}
spin-Peierls systems. Fig. \ref{fig:AlterGap}(a) inset shows the variation of
the reduced dimerization parameter $\tilde{\delta}=\delta(T)/\delta(0)$ as
function of the reduced temperature $\tau=T/T_{SP}$. Once renormalized, the dimerization parameters of
\Cl, \Br and \I present a universal thermal behavior.


Using the Barnes \etal formula \cite{Barnes1999}, the temperature dependence of $\delta$, and the exchange coupling at low temperature $J^*_{SP}$, we calculate the temperature dependence of the gap (see Fig. \ref{fig:AlterGap}(b)). Above \TSP a pseudo-gap of about 50~K is clearly visible and tend to reduce as T increases. Below \TSP the gaps open up and become temperature-independent below 0.75\TSP, reaching 180~K for \Cl and \Br and 230~K for \I.

\subsection{Electronic Interaction }\label{sec:ElecInter}

\begin{figure*} [t!]\centering
	\includegraphics[width=\linewidth]{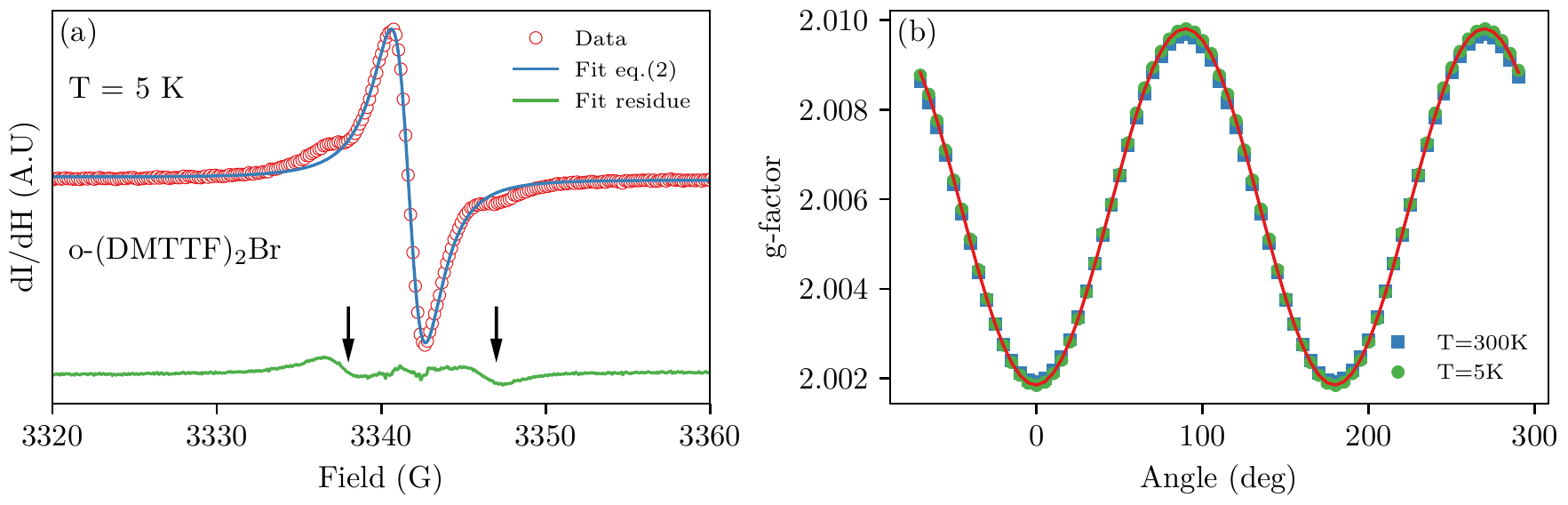} \caption{(a)
		Example of ESR signal of \Br at $T=5$~K and magnetic field $H\| c$. The central
		line has been fitted using a derivative of a lorenzian (blue line). The residue
		of the fit is presented by the green line where the arrows point the two
		satellite signals. (b) Angular dependence of the g-factor of the central line of
		defects in the spin chain measured at $T=5$~K and of the uniform spin chain line
		measured at $T=300$~K. The green line is the best fit using eq.
		\eqref{eq:g-factor_angular_dependence} with $g_\|=2.0019$ and $g_\perp=2.0098$ }
	\label{fig:Spectrum} \end{figure*}

Let us now turn to the discussion of the
low-temperature behavior. When the systems enter in the spin-Peierls phase, the
tetramerization of the DMTTF molecule stacks occurs. We have shown above that
this transition is continuous. Below about \TSP/2 \DMTTFX can be considered stabilized since the $\Delta$ and $\delta$ are
temperature independent. In a perfect and infinite system, the ground state is non-magnetic
($S=0$) and no ESR signal should be observed. However in the section \ref{sec:Susc}
we have shown the existence of a weak signal corresponding to some
10$^{-4}$.at$^{-1}$ impurities. One of the strengths of ESR is the possibility to
separate the magnetic contribution (extrinsic or intrinsic) and it was shown in
previous studies\cite{Bertaina2014a,Zeisner2019a,Soriano2020} that the ESR
signal observed at very low temperature comes from spin chain defects. By
itself, the defect is non-magnetic and is a break in the transnational symmetry
like a chain-end or a stacking fault of the alternation parameter. These
topological defects were extensively studied
theoretically\cite{Sorensen1998,Hansen1999,Nishino2000a,Fujimoto2004,Fujimoto2005} and observed by magnetometry\cite{Ami1995}, NMR \cite{Utz2017} and EPR \cite{Smirnov1998, Coulon2015, Bertaina2014a}.

ESR signal of the defects in \Br is presented in  Fig. \ref{fig:Spectrum}(a). It is recorded at $T=5$~K, which is far below \TSP. The signal contains the expected line of the spin chain defect as it was reported previously \cite{Zeisner2019,Soriano2020}, but more surprisingly, one can clearly see two shoulders on both sides of the
central line. The satellite signal is only visible in \Br. To separate and quantify the different contributions, we used to following procedure:

The blue line is the best fit using the derivative of a Lorentzian eq.
\eqref{eq:fit} of the central line. To get an accurate fit of the central line,
the points close to the shoulders have been removed from the fit procedure. Then,
we have subtracted this fit, in order to remove the contribution of the central
line from the original ESR signal. This method shows with a good accuracy the
position of the satellite lines (black arrows) since the central line does not perturb anymore the position of the satellites.
The angular dependence of the g-factor of the central line is reported in figure Fig.\ref{fig:Spectrum}(b). For comparison, the $g$-factor anisotropy of the uniform spin chain measured at
room temperature is also presented. The angular dependence of the $g$ factor can
be well described by the following relation for a $g$ tensor with uniaxial
symmetry.

\begin{equation} \label{eq:g-factor_angular_dependence} 
	g(\theta) = \sqrt{g_{\parallel}^2\cos^2\theta + g_{\perp}^2\sin^2\theta} \ \ ,
\end{equation}

Within the error limit of 10$^{-5}$ the $g$ tensors at both temperatures are identical. However, it is
important to notice that the low temperature signal cannot be attributed to the
infinite chain, which is fully dimerized at this temperature, but is rather attributed to
defects in the dimerized chain. The same angular dependence is observed in \Cl
and \I.

The presence of the satellite lines is more intriguing. 
Such a structure is often attributed to the triplet signal
\cite{Camara2010, Coulon2015} of the dimer but this should be visible at a temperature close
to the gap. However here we are at a much lower temperature and the excited state of the infinite
spin-Peierls chain must be depopulated: at $T=5$~K the relative content of spins in the first
excited state for $\Delta_{Br}=180$~K is $10^{-16}$ (see  Fig. \ref{fig:AlterGap}).
As a consequence, we can exclude that the satellites come from the excited state of the spin-Peierls chains. 

\begin{figure} \centering
	\includegraphics[width=\mycol\linewidth]{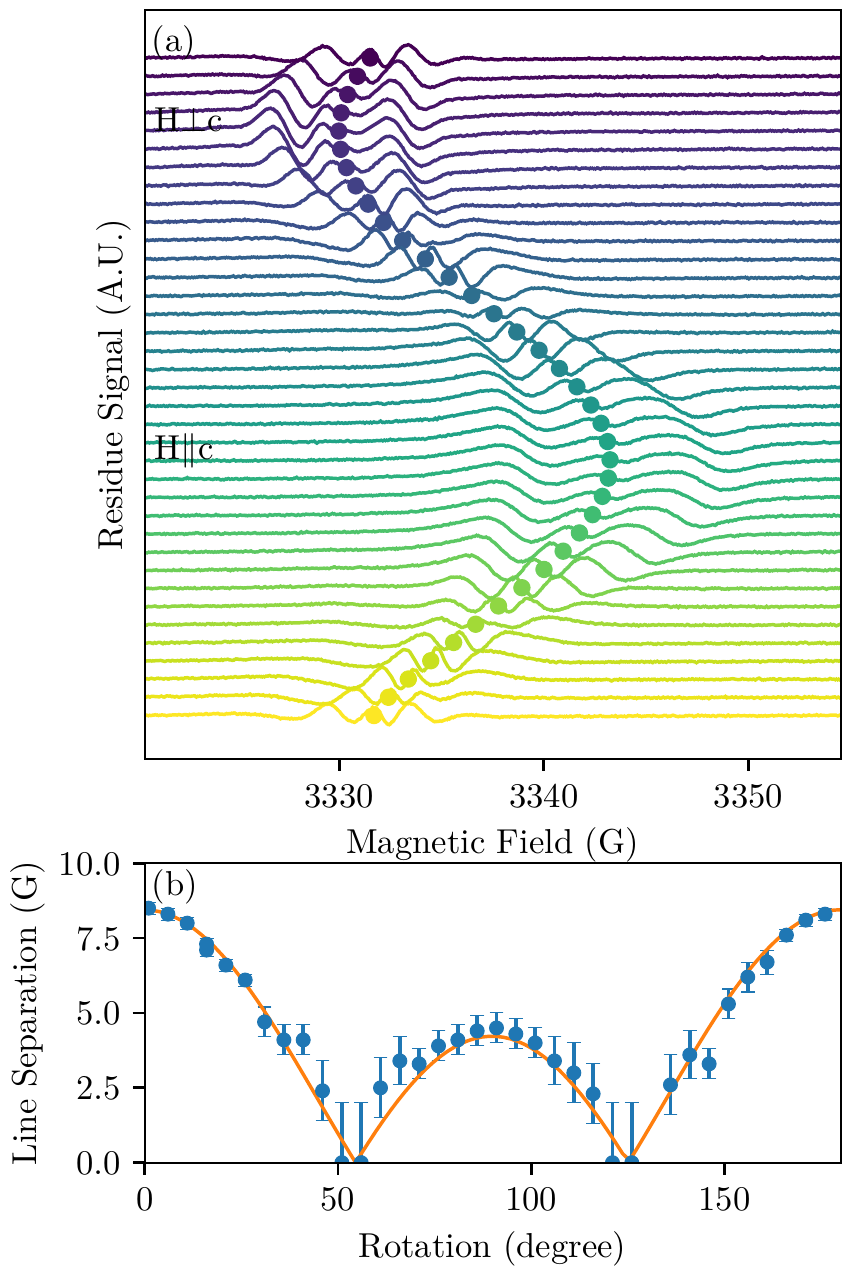}
	\caption{Angular dependence of the satellites of \Br at $T=5$~K. (a) Angular
		dependence of the fit residue. The circles show the resonance field of the central
		line. (b) Field separation of the satellites. The plain line is the best fit
		to an Ising-like expression  $d(3\cos^2\theta-1)$.}
	\label{fig:Residu} \end{figure}

The angular dependence of the satellites in \Br is presented in Fig.
\ref{fig:Residu}.  Fig. \ref{fig:Residu}(a) shows a series of fit residues
obtained by the method presented in Fig. \ref{fig:Spectrum}(a). The circles present
the resonance field of the central line. We can see that the center of gravity
of the satellites follows the angular dependence of the central
line (circles). As a consequence, the satellite signal is related to the defects of the
spin chains. Using the well resolved angular dependence of the satellite signals
of \Br, we extract the line separation as a function of the static field
orientation (Fig. \ref{fig:Residu}(b)). The error bars are due to the difficult fitting of the central line when the satellites are not clearly resolved as it is the case near the magic angle ($\sim 54^{\circ}$). The data in Fig. \ref{fig:Residu}(b) are fitted to an Ising-like
anisotropy  expression, $d(3\cos^2\theta-1)$, with $d=4.2\pm0.3$~G. It important to mention
that the satellite signal was not resolved for \Cl at any angle but the linewidth of the
central lines have shown the same anisotropy at low temperature
\cite{Zeisner2019a}. \1{Moreover the satellites were not reported in \cite{Zeisner2019a}, certainly because of saturation which prevented them for being resolved}. Intriguingly, neither the satellites nor a clear linewidth
anisotropy is observed in \I.

Such angular dependence can be attributed to a dipole-dipole interaction between
chain defects or to a $S=1$ entity like a triplon (singlet-triplet excitation)
which is split by an axial anisotropy. A pair of spins of chain defects coming
from the random distribution of the disorder is unlikely. Indeed, the
probability to find a pair of impurities with the concentration reported in Tab.
\ref{tab:impurity} is very small.  For example, using the concentration of defects
in \Br (6.6$\times 10^{-4}$) and assuming a simple cubic lattice, the
probability of finding a random pair is 3.9$\times 10^{-4}$ \cite{Behringer1958}
while the intensity of ESR signal of the satellites leads to a probability of
nearly 3 orders of magnitude bigger. Random pair defects were recently observed
by ESR on the quasi-two-dimensional organic
(BEDT-TTF)$_2$Cu[N(CN)$_2$]I \cite{Majer2020} but the concentration of defects
were substantially higher (1\%) which leads to a probability of pair existence
of 5\%.

\begin{figure} \centering \includegraphics[width=\mycol\linewidth]{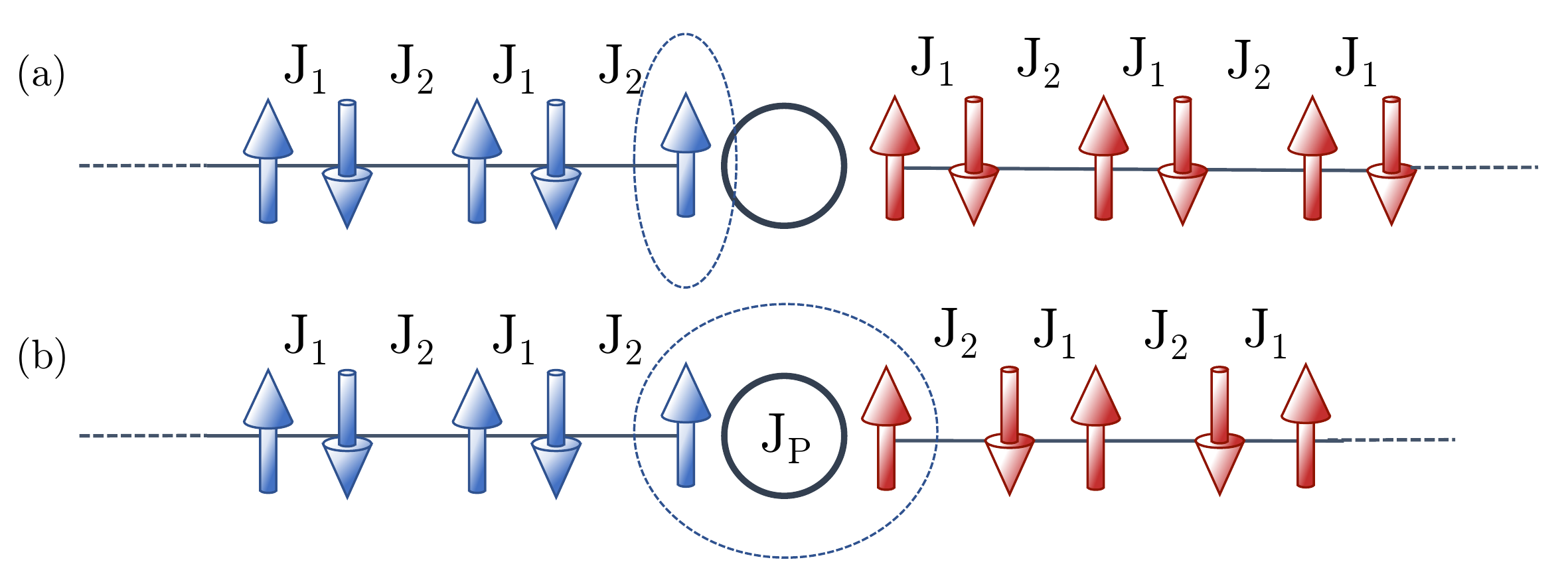}
	\caption{Schematic representation of the spin configuration around a non-magnetic
		defect in the middle of the chain. (a) When on one side of the defect
		there is an unpaired spin and on the other side the spin is paired with its
		neighbor, only the unpaired spin contributes to the ESR signal. (b) When on both
		sides of the defect the spins are unpaired, an effective coupling occurs
		leading to a triplon. } \label{fig:Model2} \end{figure}

Another explanation comes directly from the 1D nature of \DMTTFX. Fig.
\ref{fig:Model2} shows the local structure of spins induced by a break in the
translational symmetry (empty circle).
We consider only chains with a non-dimerized spin spins on the left-hand side (blue spins). On the other side of the
defect, there is also a chain (red spins). If the latter chain starts with a
strong link ($J_1=J(1+\delta)$), only the unpaired spin from the left side chain contribute
to the signal. However, if the right side chain starts with a weak link ($J_2=J(1-\delta)$)
the unpaired spins of the two finite chains are close and can interact together
with an effective coupling J$_{P}$. In this scenario, the probability of having a
pair of interacting magnetic solitons rises to 50\% of the total number of
solitons and is independent of the concentration of defects \cite{Note1}.

To prove the triplet origin of the satellite lines we performed Rabi oscillation
sequence of the ESR lines.  This pulse-ESR sequence is made of 3 pulses, le
first pulse induces a coherent rotation of the spins around the microwave field
axis and the next 2 pulses generate a Hanh echo with an intensity proportional
to the magnetization at the end of the first pulse. By adding the time dimension
to each field point of the ESR line (Fig. \ref{fig:Spectrum})(a) it is possible
to probe the nature of the spin transition even if the ESR is not
resolved\cite{Orio2021a}.

\begin{figure*} \centering \includegraphics[width=\linewidth]{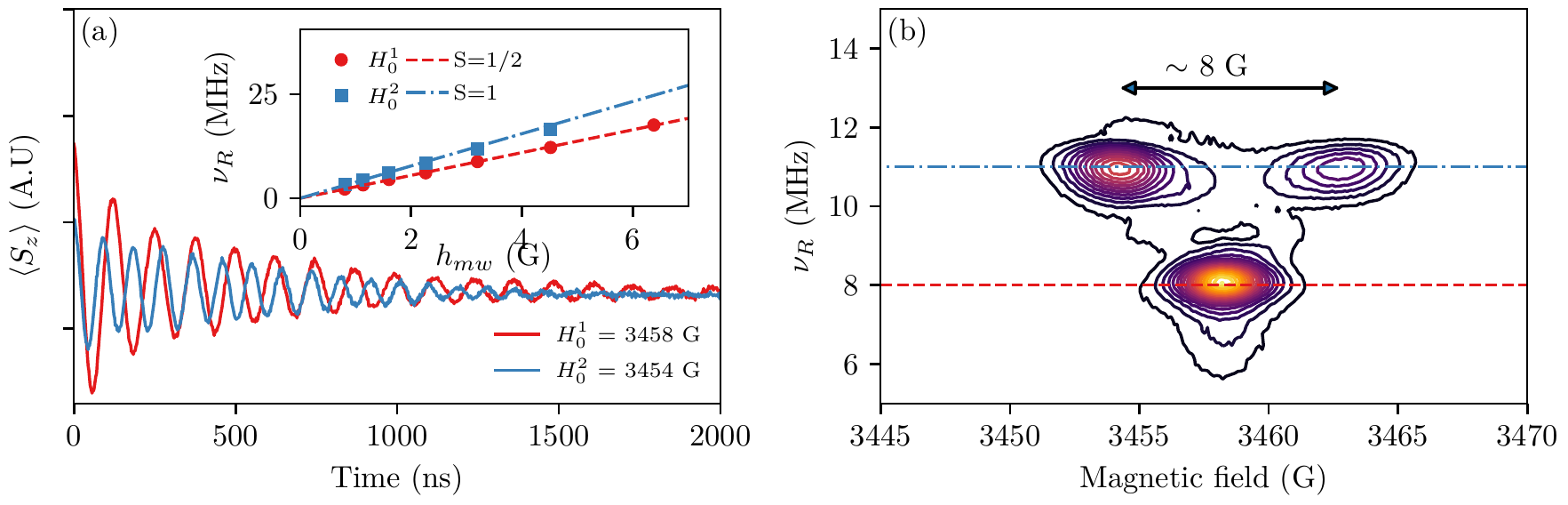}
	\caption{(a) Rabi oscillations of the strongly correlated defects of \Br at
		$T=5$~K and $H_0\|c$. The microwave frequency is fixed and the field is set at
		$H_0^1=3458$~G corresponding to the central line or $H_0^2=3454$~G to probe the
		first satellite line.  The inset is the microwave field dependence of Rabi
		frequencies within dashed red and dash dotted blue the dependence expected for 
		$S=\frac{1}{2}$ and $S=1$, respectively. (b) Contour-plot of the FFT of the
		Rabi oscillations obtained by the three-pulse sequence presented in the main
		text for $H_0\|c$.  } \label{fig:Rabi} \end{figure*}

Fig. \ref{fig:Rabi} shows the Rabi oscillations of the defect signals in \Br at $T=5$~K
and $H_0\|c$. Like in CW-ESR, the microwave frequency is fixed. Here in addition the static
field $H_0$ is fixed during the time of the sequence.  For $H_0^1=3458$~G
the central ESR line is probed, while  for  $H_0^2=3454$~G we probe one of the
satellite lines. The two lines have clearly different dynamics. For $H_0^1$
($H_0^2$), the microwave field amplitude dependence is presented in the inset
with red circles (blue squares). The dashed line is the Rabi frequency
dependence expected for a spin $S=1/2$ and the dashed dot line for the one of a
spin $S=1$ using the equation \cite{Schweiger2005}:

\begin{equation} \label{eq:RabiSpin} \nu_R^{S=1} =\sqrt{S(S+1)-S_z(S_z+1)}\times
	\nu_R^{S=1/2} \ \ \ . \end{equation}

$S_z$ is the level in which the Rabi oscillation starts, for $S=1/2$, $S_z=-1/2$ and for $S=1$, $S_z=-1$ and 0.
In the absence of a fit parameter, we found
$\nu_R(H_0^2)=\sqrt{2}\nu_R(H_0^1)$ confirming the $S=1$ nature of the satellite
lines. A field sweep Rabi oscillation sequence is presented in Fig.
\ref{fig:Rabi}(b). This figure presents the contourplot of the Rabi frequency
distribution for  $H_0\|c$ and is made of the fast Fourier transform (FFT) of
the Rabi oscillations obtained from Fig. \ref{fig:Rabi}(a) while changing the
static field. The dashed (dash doted) line shows where the Rabi frequency of a
spin $S=1/2$ ($S=1$) is expected. This method shows without ambiguity the
triplet nature of the satellite lines with a slight anisotropy $d=4$~G. In the
case of no or too weak coupling between the magnetic solitons proposed in Fig.
\ref{fig:Model2}(b) the Rabi frequency should have been the one for
$S=\frac{1}{2}$. Let us discus the cases of \Cl and \I. CW-ESR has shown no
satellite lines, no matter what the temperature and the orientation were. The field
sweep Rabi oscillation measurement shows no signature of a $S=1$ in \I but
exhibits the Rabi frequency mode of $S=1$ for \Cl (see SI). In the latter the
ESR satellite lines are unresolved in the field dimension (this is the reason why
they were undetected by CW-ESR) but are resolved in the frequency dimension (see
Supplementary Materials \cite{Note1}).This confirms the presence of a triplon state in \Cl, assumed in
Ref. \cite{Zeisner2019a}. \\
\1{It is intriguing that \I has shown no sign of paired solitons, contrary to \Br and \Cl. A possible explanation is that in \I the pinned soliton is not strictly at the end of the chain. Our model uses only explicit alternation, with no spin lattice relaxation. However, it has been shown theoretically by Hansen \etal \cite{Hansen1999} that if one takes the magneto-elastic coupling into account the magnetic soliton can be located either near the edges of the chain or can be repelled toward the interior. In the latter case no soliton pair can be formed.}

Finally, we propose an estimation of the coupling between the pairs of magnetic solitons. 

\begin{figure} \centering
	\includegraphics[width=\mycol\linewidth]{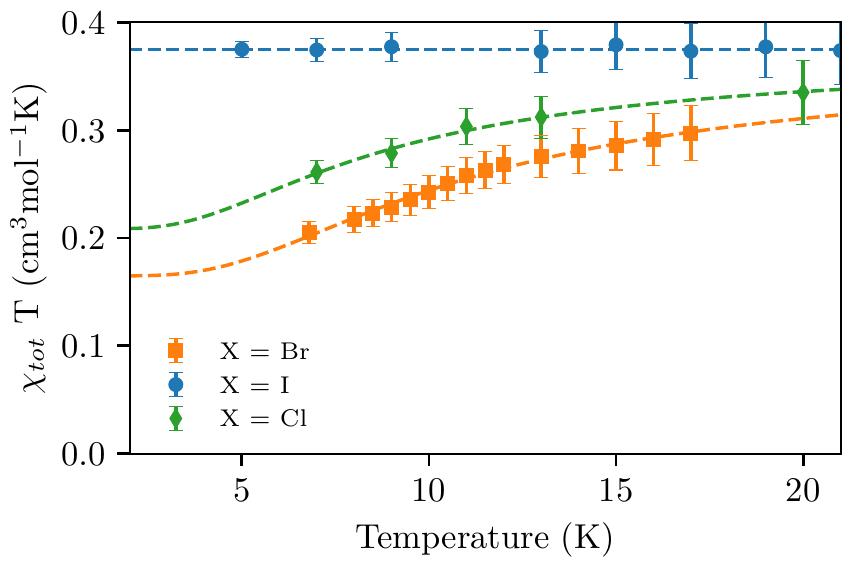}
	\caption{Product $\chi T$ as function of the temperature for the three systems $X$=Br, I and Cl, obtained from cw-ESR measurements with $H\|c$. The dashed lines are the best fits to eq. \eqref{eq:ChiT} }
	\label{fig:ChiTLT} \end{figure}

We have presented in Sec. \ref{sec:Susc} an estimation of the density of defects
based on the Curie law. A more sensitive presentation is provided by the product
of the susceptibility with the temperature $\chi T$. In the case of the
susceptibility strictly follows the Curie law, the product $\chi T$ is a
constant at any temperature (this is the case of \I in Fig. \ref{fig:ChiTLT}).
However, if some spins $S=\frac{1}{2}$ solitons form pairs with a non-negligible
coupling constant ($J_{P}$), $\chi T$ is no more temperature independent. A
naive description would be to simply considered 2 spins $S=\frac{1}{2}$ coupled
by exchange $J_{P}$. In this case the susceptibility is described by the Bleaney
and Bowers equation \cite{Bleaney1952}. However the microscopic structure and
the N-body nature of the soliton pairs presented in Fig. \ref{fig:Model2} is
more complex and necessitates DMRG and QMC calculations, briefly described in Appendic \ref{sec:A_DMRG}. It appears that the energy spectrum of a pair of solitons
is made of a singlet (S=0) ground state separated from the first excited state
(triplet S=1) by a small gap $\Delta_S$ and then the quasi continuum by a large
gap $\Delta$. If $T\ll \Delta$ we can consider only the low lying levels :
singlet-triplet. The difference with the trivial case of two coupled spins is
that the gap is no more the direct coupling between the two neighborspins but is
renormalized by the exchange couplings in the chain. QMC shows that at low
enough temperature, the susceptibility of the soliton pair can be described by
the Bleaney and Bowers formula including the gap $\Delta_S$ and DMRG calculations
show that $\Delta_S=0.35 J_{P}$.

Consequently, we fit $\chi_{tot}T$ for \Br and \Cl using a weighted sum of a single and paired solitons:

\begin{equation}\label{eq:ChiT}
	\chi T = \left[ \frac{n}{2\left(1+ \frac{1}{3}e^{\frac{\Delta_S}{kT}} \right)} + (1-n)\frac{3}{8}\right] .
\end{equation}

with $\chi T$ in emu.mol$^{-1}$K units (in number of mole of defects) $n$ the ratio of $S=\frac{1}{2}$ solitons forming
dimers and $\Delta_S$ the gap between the singlet and the first triplet (we
assume that $g=2$ for simplicity). Fig. \ref{fig:ChiTLT} shows the best fit
using $n=0.56$ and $\Delta_S=20.3$~K ($J_{P}=59.7$~K) for \Br and $n=0.44$ and
$\Delta_S=16.7$~K ($J_{P}=49.1$~K) for \Cl. The values of $n$ are coherent with
the model of a defect in the middle of the chain (see Fig. \ref{fig:Model2}) and
the effective coupling  $J_{P}$ is rather large and could lead to a
long-distance entanglement \cite{Sahling2015}. The microscopic origin of $J_{P}$
remains unclear and is beyond the scope of this article. The Ising-like
anisotropy observed in Fig. \ref{fig:Residu} can be explained by two origins:
($i$) the symmetric anisotropic exchange interaction which is consequence to the
spin orbit and can be estimated by $d=\left( \frac{\Delta g}{g}  \right)^2 J_{P}
\sim 5$~G which is in agreement with our value of $d$. 
However, it is not clear
if this formula derived for uniform superexchange interaction remains valid for
pairs of solitons.  ($ii$) The direct dipole-dipole interaction: the soliton
should not be treated as a point dipole but rather as a distribution of local magnetization
(see Appendix \ref{sec:A_DMRG}). In this way we find
$d=4.0$~G  (see Sec. \ref{sec:A_Dipol} for details) in good agreement with the experimental result.

\section{Conclusion}

In conclusion, we have presented an ESR study of \Cl, \Br and \I single crystals.
At low temperatures, these three compounds are organic gapped spin chains. In the high temperature
regime they can be treated as isotropic Heisenberg
antiferromagnetic uniform spin chains, provided that one takes into account the temperature
variation of the exchange coupling  due to the contraction of the
crystallographic cell. We have quantified the temperature dependence of the spin gap
$\Delta$ and dimerization parameter $\delta$ and shown the existence of a pseudo
gap above \TSP. Angular and temperature dependent CW-ESR measurements have
revealed the presence of magnetic solitons pinned to spin chain defects. The observation of field
sweep Rabi oscillations as well as temperature dependent ESR susceptibility
provides evidence of two different kinds of strongly correlated defects. These are, firstly, single
magnetic solitons of spin $S=\frac{1}{2}$ \1{in the three systems}. Secondly,there are pairs of exchange-coupled solitons \1{in \Br and \Cl}, whose thermally activated S=1 state (triplon) is responsible for the clearly visible second Rabi frequency. Unlike in 2D and 3D media, the density of soliton pairs in a 1D system is large - about one-half of the solitons are paired - and interdependent to the concentration of defects. As a consequence, it is
possible to coherently manipulate these quantum objects which could be of
interest in the field of quantum information processing.

\section*{Acknowledgments}
This work is supported by Agence Nationale de la Recherche (ANR project "DySCORDE",
ANR-20-CE29-0011). Financial support from the IR INFRANALYTICS FR2054 for conducting the research is gratefully acknowledged. We are grateful to C. Coulon for sharing the SQUID measurements used in
this paper.

\appendix
\section{Temperature expansion of the lattice}\label{sec:A_XRAY}

Data collection for \I was performed on an APEXII Bruker-AXS diffractometer equipped with a CCD camera and a Cryostream 700 (Oxford Cryosystems).  Sets of 3 $\omega$-scans (6$^\circ$/scan, 0.5$^\circ$/frame) were taken every 5~K, the values of the unit cell parameters used are the refined values obtained after data reduction with the Bruker SMART program.
Data collection for \Br was measured on a Rigaku Oxford Diffraction SuperNova diffractometer from 300~K down to 100~K every 50~K. In both systems the principal lattice variation is $c$, with a 2\% difference between high and low temperature, while $a$ and $b$ change as little as 0.3\% (See SI). 
\begin{figure} \centering
	\includegraphics[width=\mycol\linewidth]{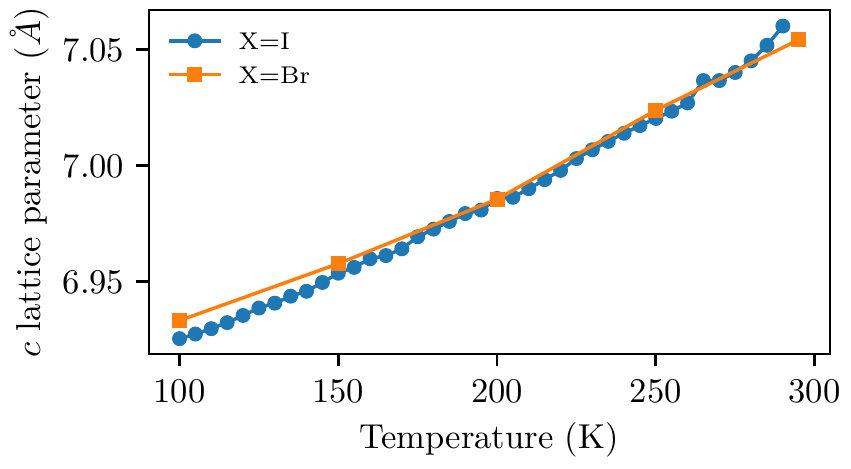}
	\caption{Temperature variation of the lattice parameter $c$ (chain direction) for \I and \Br.}
	\label{figA:XRAY} \end{figure} 

\section{DFT Calculations}\label{sec:A_DFT}

All theoretical calculations were based on the Density Functional Theory (DFT)
and were performed with the ORCA program package \cite{Neese2012}. To facilitate
comparisons between theory and experiments, X-ray crystal structure of \Br was
used. Our DFT molecular model was built considering two
dimethetyltetrathiafulvalene units together with 8 bromine counter-ions. This
model was then optimized while constraining the positions of all heavy atoms to
their experimentally derived coordinates. Only the positions of the hydrogen
atoms were relaxed because these are not reliably determined from the X-ray
structure. Geometry optimization as well as electronic structure calculations
were undertaken using the hybrid functional B3LYP \cite{Becke1993,Lee1988} in
combination with the TZV/P \cite{Schafer1994} basis set for all atoms, and by
taking advantage of the resolution of the identity (RI) approximation in the
Split-RI-J variant \cite{Weigend2006} with the appropriate Coulomb fitting sets
\cite{Klamt1993}. Increased integration grids (Grid4 and GridX4 in ORCA
convention) and tight SCF convergence criteria were used in the calculations. In
all cases, empirical dispersion corrections (D3) were included
\cite{Grimme2010}. The Heisenberg isotropic exchange coupling constants J were
evaluated from single point calculations based on the Broken Symmetry (BS)
approach\cite{Noodleman1986,Noodleman1992,Noodleman1981} using the B3LYP
functional and the TZV/P basis set. The Yamaguchi formula\cite{Soda2000} was
used to estimate the exchange coupling constants J based on the
Heisenberg–Dirac–van Vleck Hamiltonian.

We used three distances between the centers of gravity of the two (DMTTF)$_2$ molecules: 7.06\AA\xspace corresponding the X-ray value at room temperature, 6.92\AA\xspace  corresponding the X-ray value at 100~K and a contraction of 2\% (see Fig. \ref{figA:XRAY}) and 7.20\AA\xspace corresponding to a fictitious dilatation of 2\% with room temperature cell.
 
\begin{table}[h] 
	\caption{\label{tab:DFT}}
	\begin{center} 
		\begin{tabular}{lcc} \hline
			\noalign{\smallskip}\hline\noalign{\smallskip}
			 & d (\AA)  & $J_{DFT}$ (cm$^{-1}$)  \\
			\noalign{\smallskip}\hline\noalign{\smallskip} 
			RX at 300K (RT) & 7.06 & 736 \\ 
			RX at 100K (RT-2\%) & 6.92 & 648\\ 
			Fictitious (RT+2\%) & 7.20 & 832\\\noalign{\smallskip}\hline\hline 
		\end{tabular} 
	\end{center}
\end{table}

The exchange coupling values obtained by DFT overestimate the experimental values which is not surprising since we use a simple dimer model. It is more interesting to notice that a small variation of the inter molecular distance (here 2\%) induces a variation of about 13\% of $J_{DFT}$ as observed experimentally.   

\section{DMRG and QMC}\label{sec:A_DMRG}

In order to explain the electron spins interactions in the spin-Peierls phase we perform Density Matrix Renormalization Group (DMRG) and Quantum Monte Carlo  (QMC) simulations using the python ALPS toolkit \cite{Bauer2011}. We consider two magnetic structures corresponding to the model presented in Fig.\ref{fig:Model2} : (i) two dimerized 31-spin chains linked by $J_{P}$ and (ii) one dimerized 31-spin chain linked by $J_{P}$ to a dimerized 32-spin chain, represented, respectively, by $H_{SP}^{62}$ and $H_{SP}^{63}$. In all calculations we use the alternation parameter $\delta$ = 0.08 (close to the experimental value Tab. \ref{tab:Gap}) and the exchange coupling J = 1. 

\begin{multline}
	\label{hamiltonien_sp62}
	H_{SP}^{62} =\sum_{i=-1}^{-15} [J(1+\delta)S_{2i-1}.S_{2i} + J(1-\delta)S_{2i}.S_{2i+1} ] \\
		+ J_{P}S_{-1}.S_{+1} \\
	 +\sum_{i=1}^{15} [J(1-\delta)S_{2i-1}.S_{2i} + J(1+\delta)S_{2i}.S_{2i+1} ] 
\end{multline} 
\begin{multline}
	\label{hamiltonien_sp63}
		H_{SP}^{63} =\sum_{i=-1}^{-15} [J(1+\delta)S_{2i-1}.S_{2i} + J(1-\delta)S_{2i}.S_{2i+1} ] \\
		+ J_{P}S_{-1}.S_{+1}\\
	+\sum_{i=1}^{15} [J(1+\delta)S_{2i-1}.S_{2i} + J(1-\delta)S_{2i}.S_{2i+1} ] \\
	 + J(1+\delta)S_{31}.S_{32} 
\end{multline}

\begin{figure} \centering
	\includegraphics[width=\mycol\linewidth]{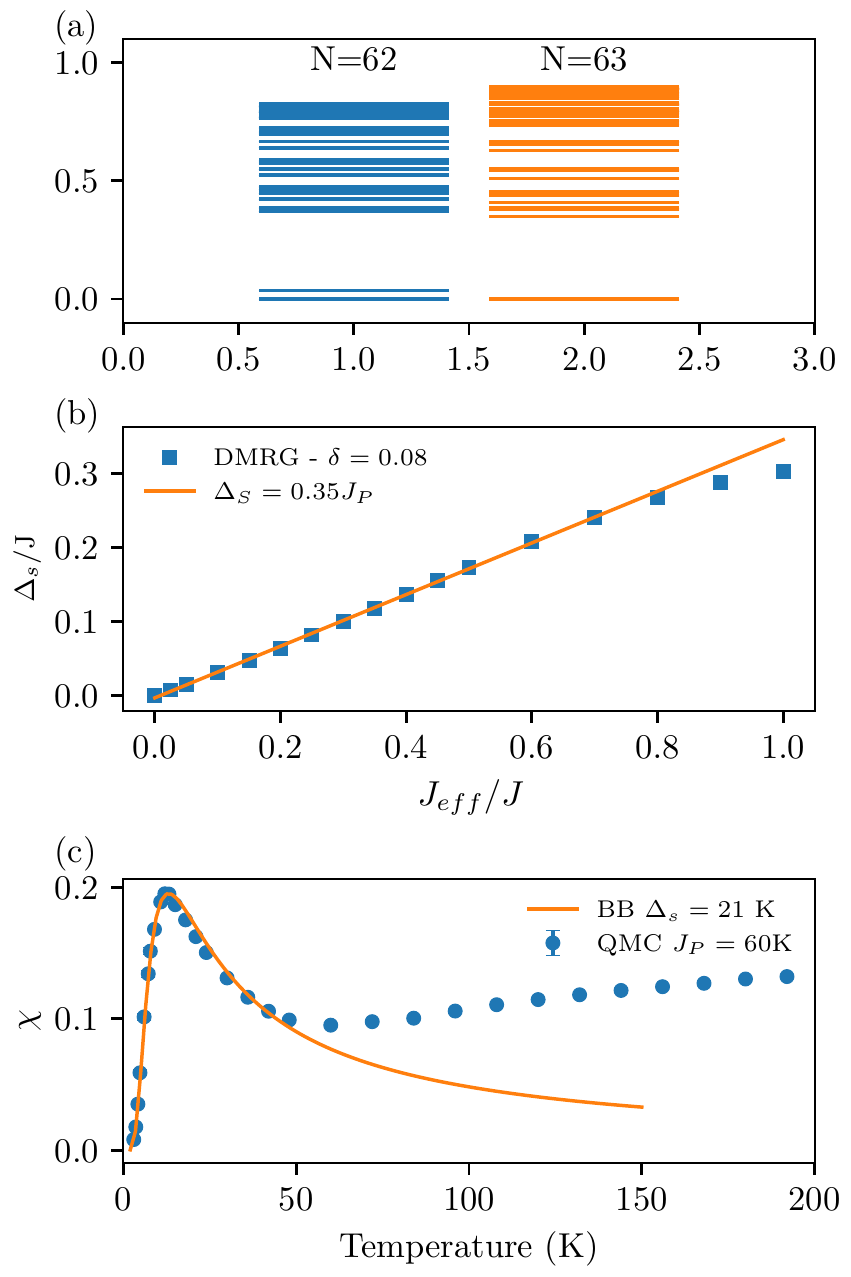} 
	\caption{(a) Energy
		spectra of a dimerized spin chain containing a defect in the middle of the chain
		(according to model in Fig. \ref{fig:Model2}). For even number of spins the
		$S_z=0$ are shown, while for odd number of spins the $S_z=\frac{1}{2}$ are
		presented. Calculation are performed by DMRG. (b) Variation of the first gap
		$\Delta_S$ of the pair of solitons between the ground state and the first
		excited state as function of the effective coupling $J_{P}$. (c) Susceptibility
		of the pair of solitons using $J_{P}=60$~K calculated by QMC. At low enough
		temperature the susceptibility can be described by the Bleaney and Bowers formula
		but with a renormalized gap (solid curve). At high temperatures the
		quasi-continuum is populated and the system cannot be treated as a pair of
		spins $\frac{1}{2}$}. \label{figA:DMRG} \end{figure}

Fig. \ref{figA:DMRG}(a) shows the first fifty eigenvalues of  $H_{SP}^{62}$ and
$H_{SP}^{63}$ calculated for $J_{P}/J=0.1$. In both cases an important gap of
energy of about $\Delta/J = 0.30$ between the ground state and the
quasi-continuum exists and this gap is directly related to the dimerization
parameter $\delta$ by $\Delta/J = 2\delta^{3/4}$. In the case N = 62 we can see a
smaller gap  of $\Delta_S/J$ = 0.035 between the ground state and the first excited one.
Fig. \ref{figA:DMRG}(b) shows the effect of $J_{P}/J$ on the gap $\Delta_S/J$.
The relation is linear and we extract a slope of 0.35.

We used the Quantum Monte Carlo method to calculate the susceptibility of
$H_{SP}^{62}$ as a function of temperature from 0.005~J to 0.5~J, which, taking $J=600$~K from experimental data (see Sec. \ref{sec:HT}) corresponds to $T=3$~K to $300$~K . For this purpose
we use the QMC algorithm "looper" which shows the best performance for
Heisenberg models. The susceptibility $\chi$ shown on the figure
\ref{figA:DMRG}(c) was calculated for $J_{P}$ = 60 K. In the low temperature
regime ($T<50$~K) the susceptibility matches with the Bleaney-Bowers formula \cite{Bleaney1952} with a
gap of $\Delta_S= 21$~K. The B-B formula is the analytical form of the
susceptibility for two spins 1/2 coupled by an isotropic exchange. Here we have
two spins 1/2 made of tens of spins coupled by $J_{P}$ and we showed by DMRG
$\Delta_S$ = 0.35$J_{P}$. The B-B formula give a fair description : 21~K/60~K =
0.35 if the gap $\Delta_S$ is renormalized by 0.35.

\section{Dipolar field}\label{sec:A_Dipol}
\begin{figure} \centering
	\includegraphics[width=\mycol\linewidth]{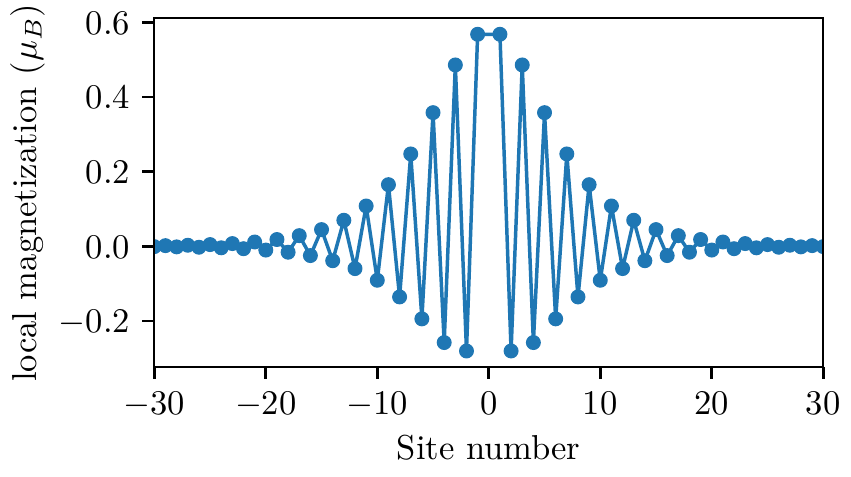}
	\caption{Local magnetization computed by DMRG of a pair of solitons in a triplet state, $S=1$, $S_z=1$ pinned to a defects in site 0. }
	\label{figA:LocMag} \end{figure} 
In order to evaluate the dipole-dipole contribution to the anisotropy
parameter $d$, we adopted a spin distribution around the defect as shown
in Fig. \ref{figA:LocMag}, all sites being equally spaced with a period of $a=7$ \AA.
The values of the magnetic moments were taken from our previous calculation
with the DMRG code of Ref. \cite{Bauer2011}:
$$
\begin{array}{lll}
	\mu_1=0.57\,\mu_{\rm B},   & ~~~~~\mu_2=-0.28\,\mu_{\rm B},  &
	~~~~~\mu_3=0.49\,\mu_{\rm B},   \\
	\mu_4=-0.26\,\mu_{\rm B},  & ~~~~~\mu_5=0.36\,\mu_{\rm B},  &
	~~~~~{\rm etc.}
\end{array}
$$
with $\mu_{-i}=\mu_i$. The dipole-dipole sum,
\begin{equation}
	d= \frac{4}{\mu_{\rm B} a^3} \sum\limits_{i=1}^N \sum\limits_{j=1}^N
	\frac{\mu_i \mu_j}{(i+j)^3}~,
	\label{eq:dipole}
\end{equation}
converges rapidly as $N\rightarrow\infty$. The final result, $d=4.0$ G,
is obtained with $N\geq5$. But even a quick estimate with $N=1$
produces a value that is only 10\% too high, $d=\mu_1^2/2\mu_{\rm B}a^3=4.4$ G.
Both theoretical estimates agree within 5\% with the value deduced from
the experimental data of Fig. 7(b), $d=4.2$ G.

\bibliography{Biblio.bib}

\pagebreak
\clearpage
\onecolumngrid

\SI

\begin{center}
	\textbf{\large{\textit{Supplementary Information} \\\smallskip
			\bluetitle{Electron spins interaction in the spin-Peierls phase of the organic spin chain \DMTTFX ($X$ = Cl, Br, I)}}}\\
	\author{L.~Soriano}\affiliation{\affIMNP}
	\author{M.D.~Kuz'min}\affiliation{\affIMNP}
	\author{O.~Pilone}\affiliation{\affIMNP}
	\author{H.~Vezin}\affiliation{\affLASIR}
	\author{O.~Jeannin}\affiliation{\affISCR}
	\author{M.~Fourmigu\'{e}}\affiliation{\affISCR}
	\author{M.~Orio}\affiliation{\affISM}
	\author{S.~Bertaina}\email{sylvain.bertaina@im2np.fr}\affiliation{\affIMNP}
\end{center}\label{Supplementary}

\section*{ESR spectra}

The data presented in the manuscript have been obtained by least-square fitting using this equation: 
\begin{equation}\label{eq:fit} 
	I_{ESR}= A\left(\frac{\Gamma\cos \phi }{\Gamma^2+(H-H_0)^2} + \frac{(H-H_0)\sin \phi}{\Gamma^2 +(H-H_0)^2} \right)
\end{equation} 

where $A$ is directly proportional to the spin susceptibility. $H_0$ is the resonance field which using the relation $g\mu_B H_0=h\nu$ give us the $g$ factor. $\Gamma$ is the half width at half maximum and $\phi$ the dispersion angle. $\Gamma$ and $\phi$ where mostly identical than the ones reported with a high accuracy in Ref. \cite{Foury-Leylekian2011}. 

\begin{figure}[!h] \centering
	\includegraphics[width=0.5\linewidth]{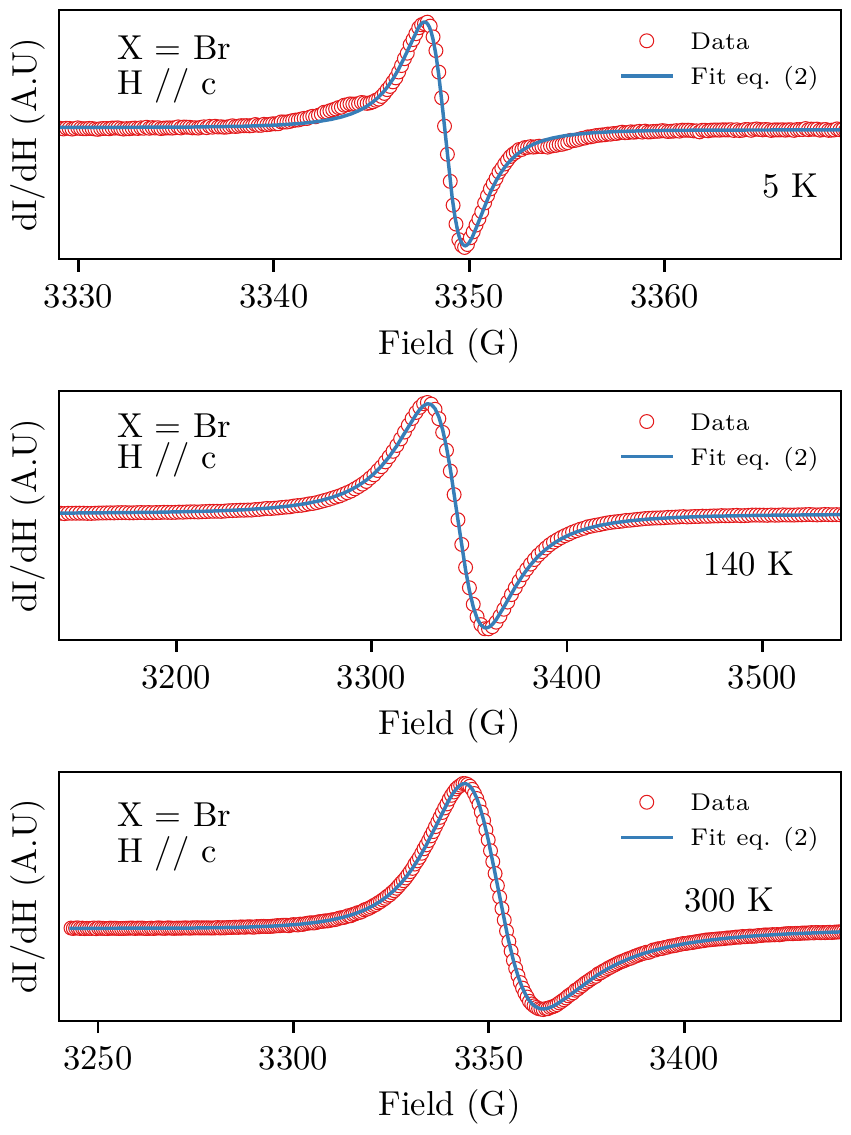}
	\caption{Examples of ESR spectra recorded for \Br at 5K, 140K and 300K. The blue line is the best fit using eq. (2) of the main text or \eqref{eq:fit}.  }
	\label{figSI_SpectraBr} 
\end{figure}
Fig. \ref{figSI_SpectraBr} present some examples of data and fit curves. Note the large variation of magnetic field scale with the temperature. 

\begin{figure}[!h] \centering
	\includegraphics[width=0.5\linewidth]{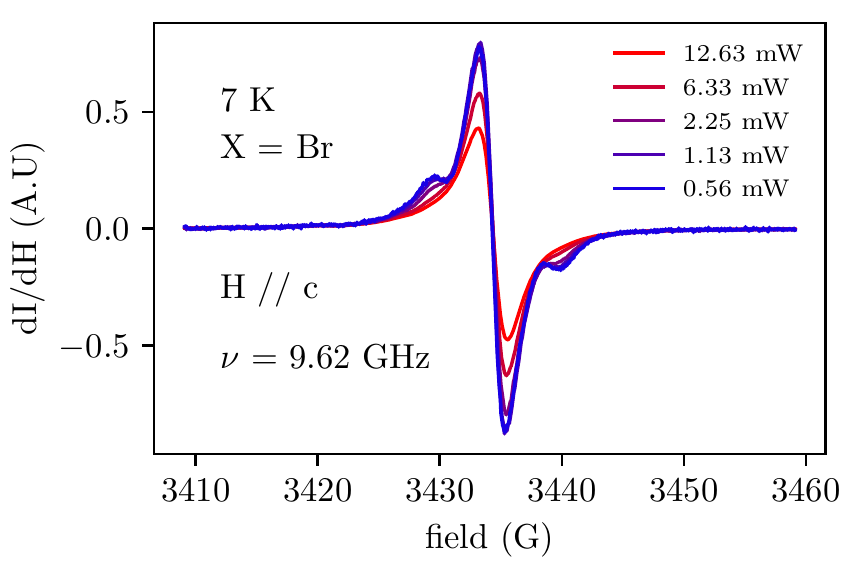}
	\caption{Microwave power dependence of ESR signal is \Br at 7K.   }
	\label{figSI_SatSatur} 
\end{figure}

It is also important noticing the difficulty to observe the satellite lines which tend to saturate easily and so can become unresolved if the microwave power is to large (see Fig \ref{figSI_SatSatur}). 

\section*{Analysis of susceptibility}

\begin{figure} \centering
	\includegraphics[width=0.5\linewidth]{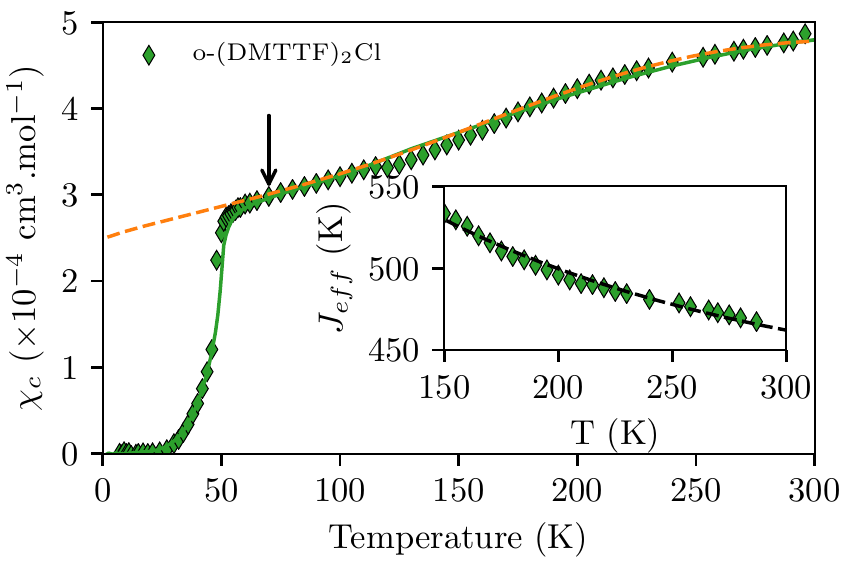}
	\caption{Temperature dependence of the spin
		susceptibility corrected by the Curie tail of $S=1/2$ defects $\chi_c$ of \Cl. The dashed lines are the theoretical values of the
		susceptibility using exchange constants $J_{eff}$ . The plain line is the susceptibility obtained by SQUID \cite{Foury-Leylekian2011} }
	\label{figSI_SuscCl} \end{figure}

Figure \ref{figSI_SuscCl}  presents the corrected susceptibility of \Cl as a function of temperature. The spin chain is uniform above $T_{SP}$ = 50 K. In this temperature range we used the method described in the section III.A.1 of main text to evaluate $J_{eff}(T)$ presented in the inset of figure \ref{figSI_SuscCl} . For $T<T_{SP}$ the Spin-Peierls transition occurs and the gapless uniform spin chains become progressively dimerized and gapped spin chains. The EPR susceptibility is in a good agreement with DC susceptibility from previous SQUID measurements \cite{Foury-Leylekian2011}.

\begin{figure*} \centering
	\includegraphics[width=\linewidth]{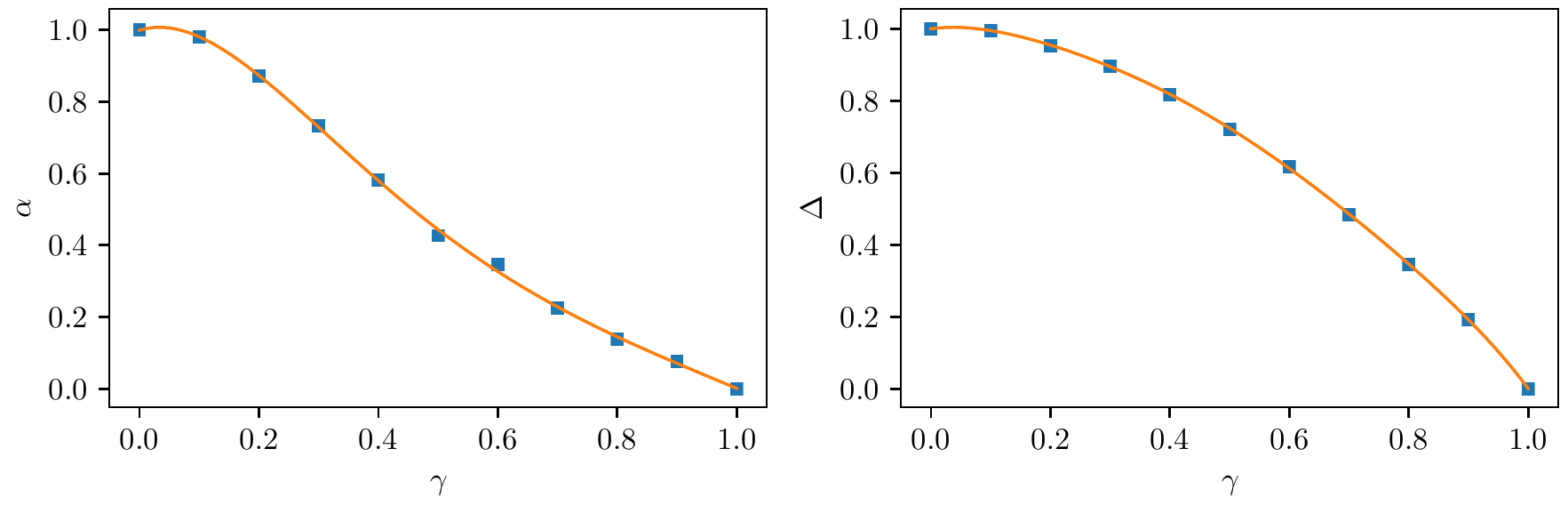}
	\caption{ Bulaevski calculation \cite{Bulaevskii1969} of a dimerized chain. The intensity $\alpha$ and gap $\Delta$ as function of $\gamma$=$\frac{1-\delta}{1+\delta}$ calculated by Bulaevskii are presented by squares. The plain lines are polynomial fit used to interpolate data to any values of $\gamma$ .}
	\label{figSI_Bull} \end{figure*}

The data from figure \ref{figSI_Bull} are adapted from the Bulaevskii \cite{Bulaevskii1969} calculations on dimerized spin chains. In the Hartree-Fock approximation and from the hamiltonian (\ref{H_Bul}) he evaluated the temperature dependences of the susceptibilities for different values of $\gamma$. At low temperature the analytic form $x(T,\gamma)$ is a good approximation with the values $\alpha(\gamma)$ and $\Delta (\gamma)$ presented in figure \ref{figSI_Bull}.
\begin{equation}
	\label{H_Bul}
	H = \sum_n S_{2n-1}.S_{2n} + \gamma S_{2n}.S_{2n+1}
\end{equation}

\begin{equation}
	\label{chi_Bul}
	x(T,\gamma) = \frac{\alpha(\gamma)}{T}e^{J_{max}\Delta (\gamma)/T}
\end{equation}

Figure \ref{figLT_Br} regroups the same data and fit parameters as in the figure 10 of the main text. On the figure \ref{figLT_Br}.b we obtain $\chi_{tot}$-$\chi_{Curie}$ by substracting the Curie law part of eq.(6) to $\chi_{tot}$. For \Cl and \Br $\chi_{tot}$-$\chi_{Curie}$ follow the characteristic form of Bowers-Bleaney \cite{Bleaney1952} model respectively with $\Delta_s$(Cl) = 16.7 K and $\Delta_s$(Br) = 20.3 K. The susceptibility $\chi_{tot}$-$\chi_{Curie}$ of \I is approximatively equal to zero in the temperature range [5 K, 25 K] and for this reason we think no S=1 gapped system exist in it.

\begin{figure*}
	
	\includegraphics[width=\linewidth]{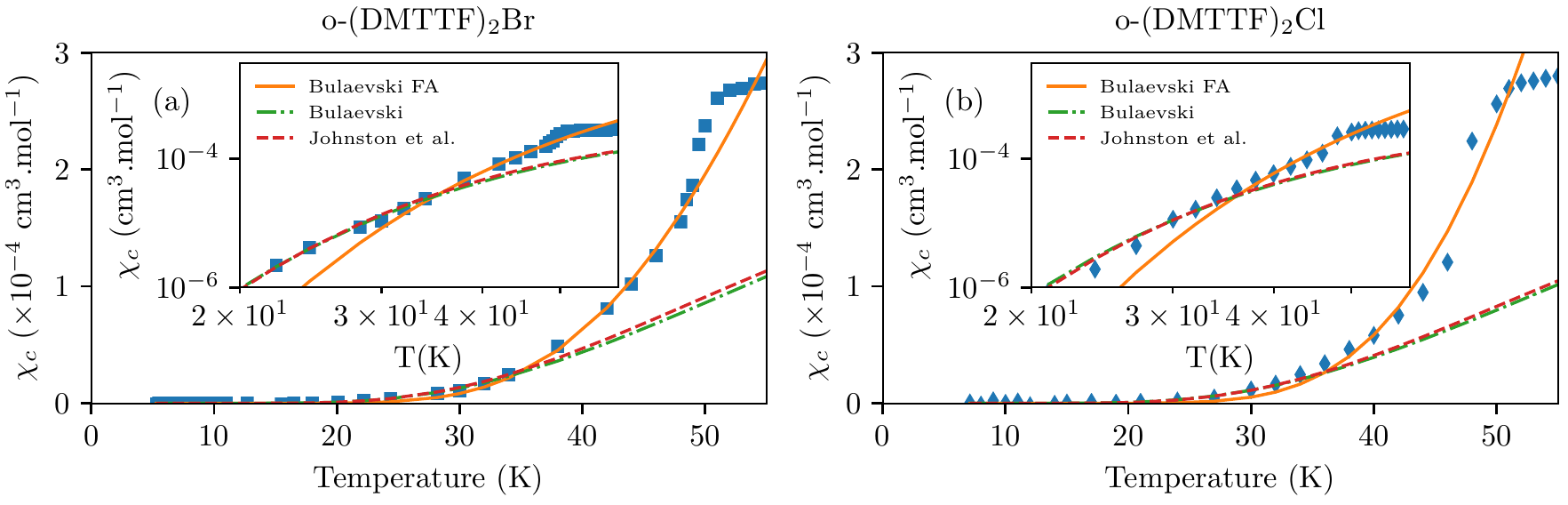}
	
	\caption{(a) Temperature dependence of the spin susceptibility in the low-temperature region for \Br. (b) Temperature dependence of the spin susceptibility in the low-temperature region for \Cl. Below $T_{SP}$ = 50 K the susceptibility drops to the non-magnetic spin-Peierls state. The plain orange lines are the best fits using eq.(3) taking $\alpha$ and $\Delta$ as two independent fit parameters. The green dashed dots and red dashed lines are the best fits for the Bulaevskii and Johnston et al. model respectively using only $\delta$ as a fit parameter. The insets are the log-log scale of the figures, magnifying the discrepancy of the Bulaevskii FA model at low temperature. }
	\label{figLT_model} 
\end{figure*}

An example of determination of the dimerization parameter $\delta$ by different methods using $\chi_{ESR}$ for \Br is given in fig.\ref{figLT_model}. The model labeled ”Bulaevski free amplitude (Bul. FA)” corresponds to eq.
(3) with $\alpha$ and $\Delta$B regarded as independent while ”Bulaevski” is eq. (3) with only $\delta$ as a free parameter. ”Johnston et al. ” is a direct numerical calculation of the
susceptibility using TMRG with only $\delta$ as a free parameter. At first sight Bul. FA seems a better fit, but a
closer look at low temperature on the log-log scale (Fig.\ref{figLT_model} inset) detects an important discrepancy with the data.
On the contrary, the two other models used, correctly show a very good agreement with experimental data for T $<$ 40 K. At higher temperature the dimerization $\delta$(T) decreases and the models cannot be used in the current form any more. Let us note a large overestimation of $\delta$ in the Bul. FA it while Bulaevski and Johnston et al. fits produce rather consistent values of $\delta$, those of Bulaevski being slightly higher.

\begin{figure*} \centering
	\includegraphics[width=1\linewidth]{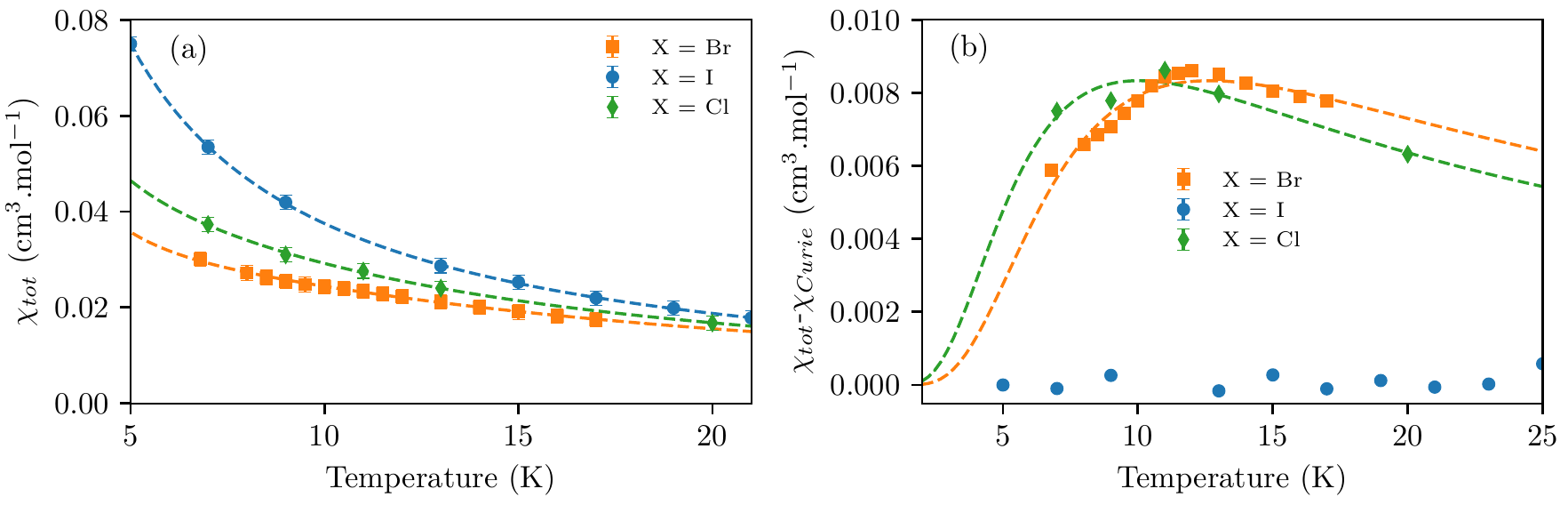}
	
	\caption{(a) Temperature dependence of the spin susceptibility in the low-temperature region for \Br, X = Cl, Br and I. The dashed lines are the best fit using eq.(6). 
		(b) Difference between the susceptibility $\chi_{tot}$ and the Curie law component in eq.(6). The dashed lines are the best fit of $\chi_{tot}$ using eq.(6) without the Curie law component.}
	\label{figLT_Br} \end{figure*}

\clearpage
\section*{Rabi field sweep}

\begin{figure} \centering
	\includegraphics[width=0.5\linewidth]{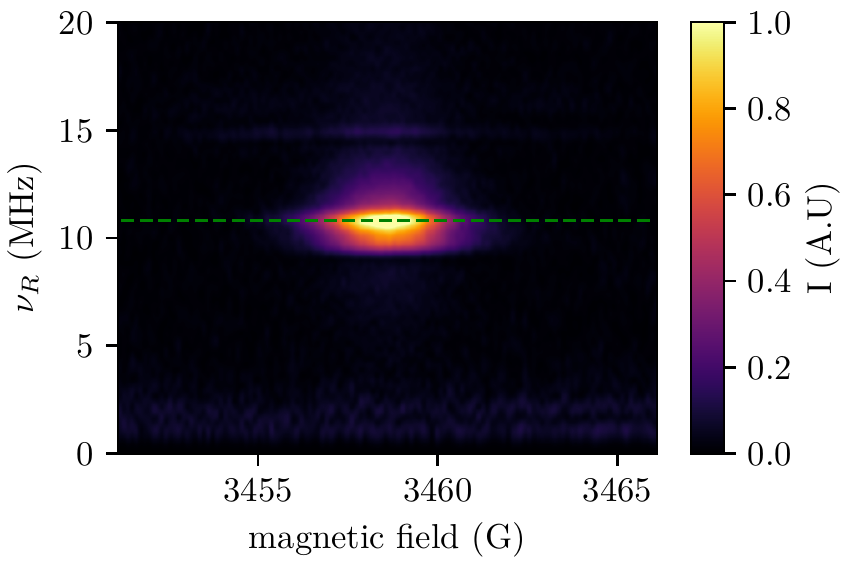}
	\caption{Image plot of the FFT of the Rabi oscillations of \I for $H_0$ // c, T = 5.8 K and $h_{mw}$ = 4 G.We used the three-pulse sequence described in experimental details.}
	\label{figSI_Rabi2DI} \end{figure}

\begin{figure} \centering
	\includegraphics[width=0.5\linewidth]{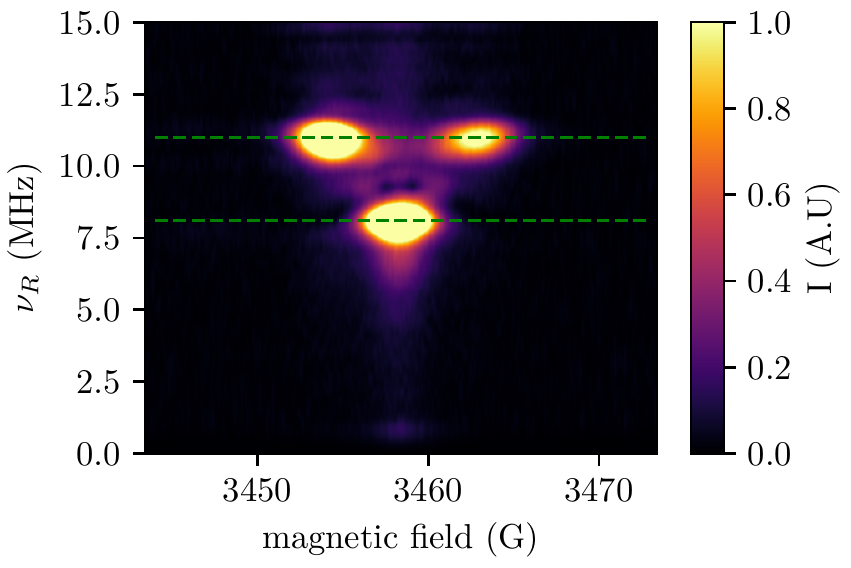}
	\caption{Image plot of the FFT of the Rabi oscillations of \Br for $H_0$ // c, T = 5.4 K and $h_{mw}$ = 2.7 G.We used the three-pulse sequence described in experimental details.}
	\label{figSI_Rabi2DBr} \end{figure}

\begin{figure} \centering
	\includegraphics[width=0.5\linewidth]{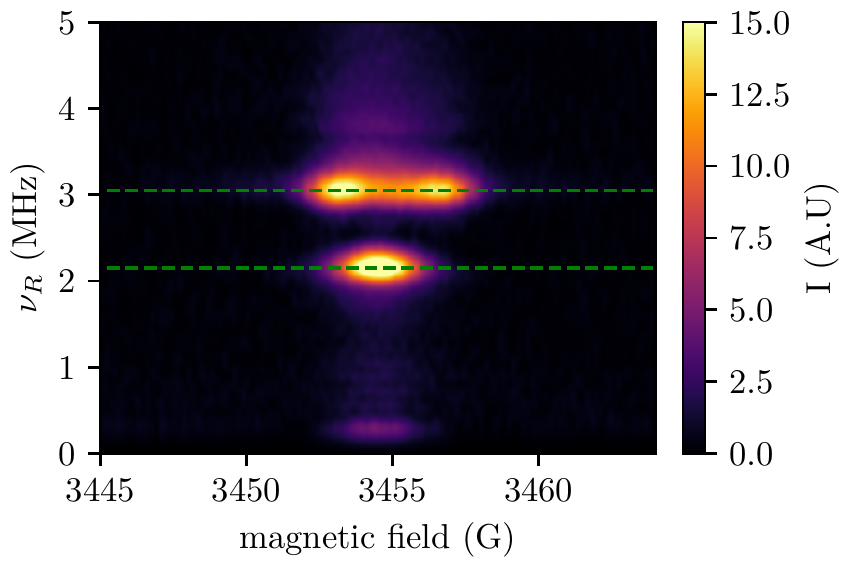}
	\caption{Image plot of the FFT of the Rabi oscillations of \Cl for $H_0$ //, T = 5.5 K and $h_{mw}$ = 0.75 G .We used the three-pulse sequence described in experimental details.}
	\label{figSI_Rabi2DCl} \end{figure}

The figure \ref{figSI_Rabi2DI},\ref{figSI_Rabi2DBr} and \ref{figSI_Rabi2DCl}  show the FFT of the Rabi oscillations on \DMTTFX (X = Cl, Br and I) sweeping the magnetic field. We can clearly identify two different Rabi frequencies at $\nu_R^{S=1/2}$ = 7.6 MHz and $\nu_R^{S=1}$ = 11 MHz in \Br. The two spots attributed to S = 1 are separated by $\sim$ 8 G. On \Cl  the S= 1/2 system oscillates at $\nu_R^{S=1/2}$ = 6.1 MHz and one can see a faint large spot at $\nu_R^{S=1}$ = 8.5 MHz . The figure \ref{figSI_Rabi2DI} confirms that only S = 1/2 exist in \I only a spot at  $\nu_R^{S=1/2}$ = 11 MHz is visible. 

\newpage
\section*{EPR Spectra, fit and residue}

\begin{figure*} \centering
	\includegraphics[width=\linewidth]{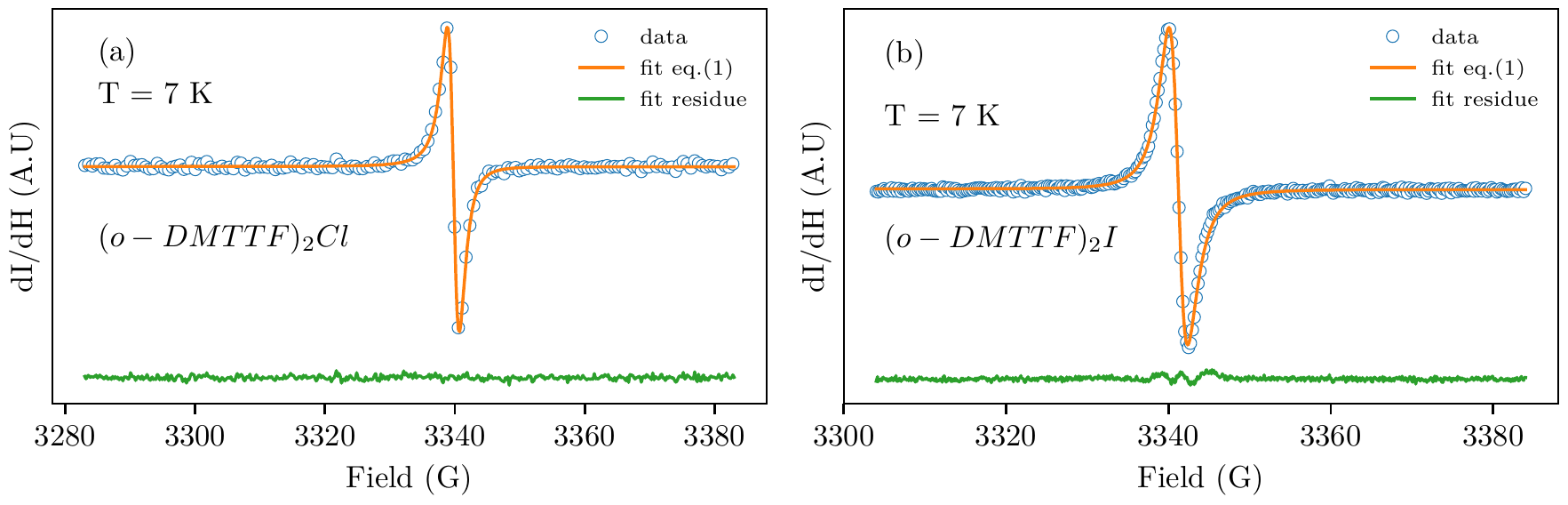}
	\caption{(a) Example of ESR signal of for\Cl and (b) \I at T = 7 K and magnetic field H // c. The central lines have been fitted using a derivative of a lorentzian (orange lines). The residues of the fits are represented by the green line. }
	\label{fig_spectra} \end{figure*}

On the figure \ref{fig_spectra} the spectra of \Cl and \I are fitted with a dispersive lorentzian model are presented. This fit is really suitable for \Cl for  every orientations and gives a very small residue. The \I residue is bigger because the line is not exactly lorentzian.By looking at the angular dependence ... on the figure 7 of the main text.

\begin{figure*} \centering
	\includegraphics[width=\linewidth]{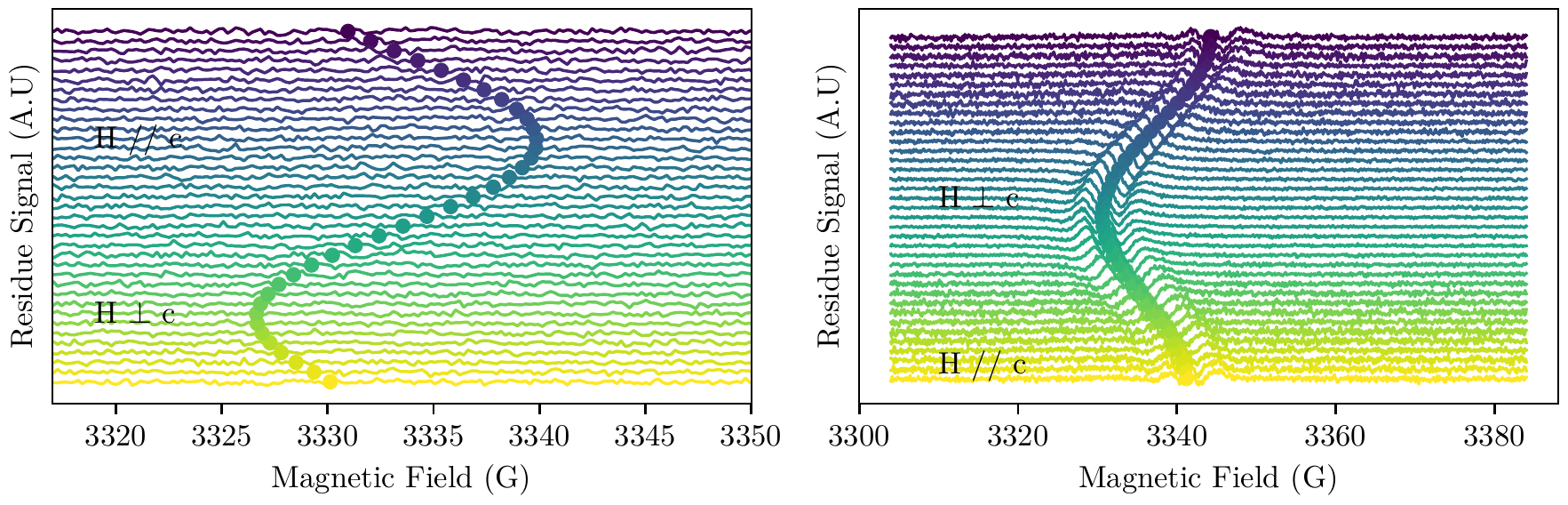}
	\caption{Angular dependence of the fit
		residue of \Cl (left) and \I (right) at 7 K. The circles show the resonance field of the central line.  }
	\label{fig_residue} \end{figure*}

\section*{Probability of paired and single solitons}
In the main text we have provide an explanation of 50/50\% change for having a paired or a single soliton. Here we list all the possible configurations with:  odd number of spins with the soliton on the right $O_R$ of on the left $O_L$ and the even number of spins with strong links on the edge $E_S$ or weak links $E_W$. Only the result around the defect is considered. 8 configurations give a single soliton and 4 give  paired solitons
leading to 50/50\%
\begin{table}[h] \caption{List of the configurations. The defect is between the "left chain" and the "right chain". Around  the defect, the spin chain ca be odd number of spins with the non dimerized spin on the left $O_L$ or the right $O_R$, or even number of spins with strong link $E_S$ or weak link $E_W$ on the edge.   }
	\begin{center} \begin{tabular}{ccc} \hline
			\noalign{\smallskip}\hline\noalign{\smallskip} 
			Left chain &  Right chain & Nature  \\
			\noalign{\smallskip}\hline\noalign{\smallskip} 
			$E_W$ & $E_W$ & Paired \\ 
			$E_W$& $E_S$ & Single\\
			$E_W$& $O_L$ & Paired\\
			$E_W$&  $O_R$&  Single \\
			$E_S$ & $E_W$ & Single\\ 
			$E_S$& $E_S$ & None\\
			$E_S$& $O_L$ & Single\\
			$E_S$&  $O_R$&  None\\
			$O_L$ & $E_W$ & Single \\ 
			$O_L$& $E_S$ & None\\
			$O_L$& $O_L$ & Single\\
			$O_L$&  $O_R$&  None \\
			$O_R$ & $E_W$ & Paired\\ 
			$O_R$& $E_S$ & Single\\
			$O_R$& $O_L$ & Paired\\
			$O_R$&  $O_R$&  Single\\

			\noalign{\smallskip}\hline\hline \end{tabular} \end{center}
\end{table}  

\clearpage
\section*{Python code for Johnston \etal susceptibility}
\begin{python}
	#!/usr/bin/python
	# -*- coding: utf-8 -*-
	
	import numpy as np
	
	def chi_J(t, a):
	""" Chi_star as a function of temperature according to 
	Johnston et al. (2000) p. 9578 for S = 1/2 HAF chain
	t ... reduced temperature (k_B*T/Jmax)
	a ... asymmetry parameter alpha = (J2/J1) =(1-delta)/(1+delta)
	values for Nmn and Dmn parameters are taken from Table II in Johnston et al.
	(2000) """
	
	delta_fit = 1. - 0.5 * a - 2. * a ** 2 + 1.5 * a ** 3
	g1 = 0.38658545
	g2 = -0.20727806
	delta_0 = (1. - a) ** 0.75 * (1. + a) ** 0.25 + g1 * a * (1. - a) \
	+ g2 * a ** 2 * (1. - a) ** 2
	N0 = 1.
	N1 = 0.63427990 - 2.06777217 * a - 0.70972219 * a ** 2 + 4.89720885 \
	* a ** 3 - 2.80783223 * a ** 4
	N2 = 0.18776962 - 2.84847225 * a + 5.96899688 * a ** 2 - 3.85145137 \
	* a ** 3 + 0.64055849 * a ** 4
	N3 = 0.033603617 - 0.757981757 * a + 4.137970390 * a ** 2 \
	- 6.100241386 * a ** 3 + 2.701116573 * a ** 4
	N4 = 0.0038611069 + 0.5750352896 * a - 2.3359243110 * a ** 2 \
	+ 2.934083364 * a ** 3 - 1.1756629304 * a ** 4
	N5 = 0.00027331430 - 0.10724895512 * a + 0.40345647304 * a ** 2 \
	- 0.48608843641 * a ** 3 + 0.18972153852 * a ** 4
	N6 = 0.00578123759 * a - 0.02313572892 * a ** 2 + 0.02892774508 * a \
	** 3 - 0.01157325374 * a ** 4
	N71 = 2.59870347E-7
	N72 = -2.39236193E-7
	sum_Nn = N0 + N1 / t + N2 / t ** 2 + N3 / t ** 3 + N4 / t ** 4 + N5 \
	/ t ** 5 + N6 / t ** 6
	D0 = 1.
	D1 = -0.11572010 - 1.31777217 * a + 1.29027781 * a ** 2 \
	+ 3.39720885 * a ** 3 - 2.80783223 * a ** 4
	D2 = 0.08705969 - 1.44693321 * a + 5.09401919 * a ** 2 \
	- 10.51861382 * a ** 3 + 8.97655318 * a ** 4 + 5.75312680 * a \
	** 5 - 11.83647774 * a ** 6 + 4.21174835 * a ** 7
	D3 = 0.00563137 + 0.65986015 * a - 1.38069533 * a ** 2 - 0.09849603 \
	* a ** 3 + 7.54214913 * a ** 4 - 22.31810507 * a ** 5 \
	+ 27.60773633 * a ** 6 - 6.39966673 * a ** 7 - 15.69691721 * a \
	** 8 + 13.37035665 * a ** 9 - 3.15881126 * a ** 10
	D4 = 0.0010408866 + 0.1008789796 * a - 0.9188446197 * a ** 2 \
	+ 1.6052570070 * a ** 3 - 0.7511481272 * a ** 4
	D5 = 0.0000683286 - 0.1410232710 * a + 0.6939435034 * a ** 2 \
	- 0.9608700949 * a ** 3 + 0.4106951428 * a ** 4
	D6 = 0.0367159872 * a - 0.1540749976 * a ** 2 + 0.1982667100 * a \
	** 3 - 0.0806430233 * a ** 4
	D7 = -0.00314381636 * a + 0.01140642324 * a ** 2 - 0.01338139741 \
	* a ** 3 + 0.00511879053 * a ** 4
	D81 = 1.25124679E-7
	D82 = -1.03824523E-7
	sum_Dn = D0 + D1 * t ** -1 + D2 * t ** -2 + D3 * t ** -3 + D4 * t \
	** -4 + D5 * t ** -5 + D6 * t ** -6 + D7 * t ** -7
	y = 4.69918784
	z = 3.55692695
	Pade_approx = (sum_Nn + (N71 * a + N72 * a ** 2) * (delta_0 / t)
	** y * t ** -7) / (sum_Dn + (D81 * a + D82 * a ** 2)
	* (delta_0 / t) ** z * np.exp((delta_0 - delta_fit) / t)
	* t ** -8)
	chi_star = Pade_approx * np.exp(-delta_fit / t) / (4 * t)
	return 0.3751 * 2.01 ** 2 / J * chi_star

\end{python}

\end{document}